\documentclass[prd,aps,10pt,twocolumn,superscriptaddress,notitlepage,nofootinbib,nobibnotes,amssymb,amsmath]{revtex4-1}
\pdfoutput=1
\usepackage[utf8]{inputenc}
\usepackage[T1]{fontenc}
\usepackage{lmodern, textcomp}
\usepackage[british]{babel}
\usepackage{csquotes}

\usepackage{eurosym, xspace}
\usepackage{graphicx, float}
\usepackage{subcaption}

\usepackage[font=small,labelfont=bf,justification=justified, format=plain]{caption}

\usepackage{amsmath, amsfonts, amssymb}
\usepackage{mathtools}
\usepackage{siunitx}

\usepackage{xcolor}

\newcommand{\beqn}{\begin{eqnarray}}
\newcommand{\eeqn}{\end{eqnarray}}
\newcommand{\eq}[1]{(\ref{#1})}

\newcommand{\bs}{\boldsymbol}

\newcommand{\avr}[1]{{\left\langle #1 \right\rangle}}

\newcommand{\beqs}{\begin{subequations}}
\newcommand{\eeqs}{\end{subequations}\\[-3.5mm]\noindent}

\newcommand{\Z}[0]{\ensuremath{\mathbb{Z}}}

\renewcommand{\Re}{\operatorname{Re}}
\renewcommand{\Im}{\operatorname{Im}}

\newcommand{\pd}[0]{\partial}

\newcommand{\mean}[1]{\langle #1 \rangle}

\newcommand{\abs}[1]{{|#1|}}

\DeclareMathOperator{\argmax}{argmax}
\DeclareMathOperator{\Var}{Var}

\usepackage{marginnote, xparse}
\newcounter{draftcommentcnt}
\NewDocumentCommand{\draftcomment}{s O{red} m}{%
	\def\margnote{\IfBooleanTF{#1}{\marginnote}{\marginpar}}%
	\stepcounter{draftcommentcnt}%
	\textcolor{#2}{#3}%
	\margnote{\textcolor{#2}{$\Leftarrow$ \arabic{draftcommentcnt}}}%
}

\begin{document}

\title{Topological defects and confinement with machine learning: \\ the case of monopoles in compact electrodynamics}

\author{M. N. Chernodub}
\affiliation{Institut Denis Poisson CNRS/UMR 7013, Universit\'e de Tours, 37200 France}
\affiliation{Pacific Quantum Center, Far Eastern Federal University, Sukhanova 8, Vladivostok, 690950, Russia}
\author{Harold Erbin}
\affiliation{Dipartimento di Fisica, Universit\'a di Torino, INFN Sezione di Torino and Arnold-Regge Center, Via Pietro Giuria 1, I-10125 Torino, Italy}
\author{V. A. Goy}
\affiliation{Pacific Quantum Center, Far Eastern Federal University, Sukhanova 8, Vladivostok, 690950, Russia}
\author{A. V. Molochkov}
\affiliation{Pacific Quantum Center, Far Eastern Federal University, Sukhanova 8, Vladivostok, 690950, Russia}

\begin{abstract}
We investigate the advantages of machine learning techniques to recognize the dynamics of topological objects in quantum field theories. We consider the compact U(1) gauge theory in three spacetime dimensions as the simplest example of a theory that exhibits confinement and mass gap phenomena generated by monopoles. We train a neural network with a generated set of monopole configurations to distinguish between confinement and deconfinement phases, from which it is possible to determine the deconfinement transition point, and to predict several observables. The model uses a supervised learning approach and treats the monopole configurations as three-dimensional images (holograms). We show that the model can determine the transition temperature with accuracy, which depends on the criteria implemented in the algorithm. More importantly, we train the neural network with configurations from a single lattice size before making predictions for configurations from other lattice sizes, from which a reliable estimation of the critical temperatures are obtained.
\end{abstract}

\date{June 16, 2020}

\maketitle

\section{Introduction}

Compact Abelian gauge model in two spatial dimensions mimics several exciting nonperturbative features of Quantum Chromodynamics (QCD), including the linear confinement of electric charges at large distances and mass-gap generation~\cite{Polyakov:1976fu}. This Abelian toy model -- often called compact Electrodynamics, or cQED -- possesses topologically stable objects, monopoles, which reveal themselves as instantons. The instantons also appear in a Euclidean formulation of QCD~\cite{Gross:1980br}, thus bringing an additional bridge between these two theories. In the presence of fermions, the monopoles catalyze the chiral symmetry breaking in cQED~\cite{Fiebig:1990uh}. The chiral symmetry plays a very important role in the hadronic physics described by QCD. Finally, both QCD and cQED experience a finite-temperature transition to a high-temperature phase that lacks the linear confinement property.  

In addition to its role in particle physics, cQED also serves as a useful macroscopic model in a broad class of condensed matter systems~\cite{ref:book:Herbut,ref:book:Kleinert}. It experiences mesoscopic phenomena like the Casimir effect~\cite{Chernodub:2017mhi}, which mimics closely its non-Abelian analog~\cite{Chernodub:2018pmt}, and may be explored with machine learning techniques~\cite{Chernodub:2019kon} similar to the ones discussed in this paper.

Contrary to the theory of strong interactions, the nonperturbative effects in cQED are well understood. The confinement and mass gap generation admit an analytical treatment in a weak coupling regime of zero-temperature cQED~\cite{Polyakov:1976fu}, while the phase transition may be characterized both by analytical and numerical techniques~\cite{Parga:1981tm,Coddington:1986jk,Chernodub:2001ws,Borisenko:2008sc,Borisenko:2010qe,Caselle:2019khe}. In the context of Abelian theory, the dynamics of the Abelian monopoles can explain all these non-perturbative phenomena. 

In our paper, we consider the finite-temperature phase transition in cQED and the associated monopole dynamics using the machine learning (ML) approach. ML techniques stand on powerful programming tools that allow a computer to find a way to perform a certain task without being explicitly preprogrammed in advance (we refer the interested reader to Ref.~\cite{ref:review:1,ref:review:2} for physicist reviews). In the approach we intend to use, a neural network is trained to compute some target features from a given configuration by providing a certain number of examples. Then, the network can be used to predict the target variables for any configuration both inside and outside the domain of training.
In other words, the neural network learns how to predict a required feature of a complex system and then uses the acquired knowledge to make the predictions independently. Given the impressive versatility of the approach, ML methods find their implementations in studies of the phase structure of various many-body systems, strongly-correlated environments, and field theories~\cite{ref:phases:0,ref:phases:1,ref:phases:2,ref:phases:3,ref:phases:4, ref:phases:6,ref:phases:7,ref:phases:8,ref:phases:9,ref:phases:10, ref:phases:11, ref:phases:12,Bachtis:2020dmf}.

The use of the ML techniques has various motivations. Evidently, the neural networks offer a clear computational advantage: while the learning phase of the neural networks may be slow,\footnote{The slowness of the learning phase is not a necessity. For example, the neural network of this paper trains in few minutes.} their predictions are usually coming very fast. Therefore, ML methods became particularly successful in the investigation of many complex physical systems that involve a high number of degrees of freedom where the traditional methods provide slow advance.

ML approaches are also believed to be useful for uncovering hidden mechanisms of physical phenomena that otherwise lack a solid theoretical explanation. In the first stage, the neural network learns the effect in question in the system with many (infinite in the thermodynamic limit) degrees of freedom. Then the ML algorithm demonstrates the successful implementation of the learning phase by recognizing the phenomenon at new (to the algorithm) configurations of the same system. The successful completion of the examination phase implies that a finite-element neural network has managed to successfully describe the system with a vast number of degrees of freedom. Thus, the third stage consists of learning what the neural network learned during the training stage about the phenomenon in question by analyzing the weights of the neurons inside the network. This procedure may give an insight on the mechanism of the physical effect as it was learned by the neural network.This third step is outside the scope of our paper and we will focus on demonstrating that a neural network can learn to compute the quantities we are interested in.

Our paper aims to investigate how well the ML techniques may see the deconfining phase transition in a field theory through the eyes of topological defects. We use the compact electrodynamics in which the finite-temperature phase transition is tightly related with the dynamics of the monopoles. The lattice formulation of cQED allows for a straightforward identification of monopoles while the imprint of the phase transition on the monopole dynamics is well known: the system goes from the monopole gas at low temperature to a gas of monopole-antimonopole pairs at high temperature through a phase transition of the Berezinskii--Kosterlitz--Thouless type~\cite{Berezinskii:1971fst,Berezinskii:1971scn,Kosterlitz:1973xp}.\footnote{Notice that an inclusion of the matter fields may shift the location and change the type of the finite temperature deconfining phase transition~\cite{Dunne:2000vp}.}

A particular question we address in this paper is whether neural network can extrapolate predictions for configurations at different lattice sizes after having been trained with configurations from a single lattice size. We will see that it is indeed the case, implying that the neural network automatically captures the notion of the thermodynamic limit. While the quantities predicted for individual configuration are not particularly accurate, we find that the neural network still understands that something happens to the system; The acquired knowledge allows the neural network to determine the critical temperature to a good accuracy.

We provide a basic description of the compact electrodynamics on the lattice, the lattice monopoles, and the relevant observables in Section~\ref{sec:model}. The neural network used in our analysis appears in Section~\ref{sec:neural} while Section~\ref{sec:results} represents the results of the application of the machine learning methods to the monopole configurations. The last Section summarizes our conclusions. 

\section{Gauge model}
\label{sec:model}

The term ``compact electrodynamics'' describes an Abelian U(1) gauge model, which admits the existence of the monopole-like singularities in the gauge field. We consider a lattice version of this model because the lattice regularization offers the most natural way to describe the compact gauge fields. We study the Wick-rotated model in three Euclidean space-time dimensions because we are interested in thermal equilibrium states, which can be studied numerically in the Euclidean version of the model.

\subsection{Compact electrodynamics on the lattice}

The compact lattice electrodynamics is described by the following action:
\beqn
S[\theta] = \beta \sum_P \left(1 - \cos \theta_P \right)\,.
\label{eq:S}
\eeqn
where the sum runs over all elementary plaquettes $P \equiv P_{x,\mu\nu}$ of the lattice. Each plaquette $P_{x,\mu\nu}$ is labeled by the position~$x$ of one of its corners and by two orthogonal vectors $\mu < \nu$ that determine the orientation of plaquette in the Euclidean spacetime ($\mu,\nu = 1, 2, 3$).

The plaquette angles in the action~\eq{eq:S}
\beqn
\theta_{P_{x,\mu\nu}} = \theta_{x,\mu} + \theta_{x+\hat\mu,\nu} - \theta_{x+\hat\nu,\mu} - \theta_{x,\nu}\,,
\label{eq:theta:P}
\eeqn
play a role of the lattice field strength of the compact gauge field $\theta_{x,\mu} \in [-\pi,+\pi)$. 
This dimensionless compact variable has a vector nature: the field $\theta_{x,\mu}$ is defined at the link starting at the point $x$ and pointing in the direction $\mu$.

The lattice angle $\theta_{x,\mu} = a A_{\mu}(x)$ is the dimensionless suitable for numerical simulations. It is related the continuum gauge field $A_\mu(x)$ via the length of an elementary lattice link~$a$. In the continuum limit, the lattice spacing tends to zero, $a\to 0$, and the plaquette variable~\eq{eq:theta:P} approaches its continuum expression 
\beqn
\theta_{P_{x,\mu\nu}} = a^2 F_{\mu\nu}(x) + O(a^4),
\label{eq:theta:cont}
\eeqn
where $F_{\mu\nu} = \partial_\mu A_\nu - \partial_\nu A_\mu$ is the field strength tensor in the continuous spacetime. The validity of this identification is constrained by the absence of singular monopole-like configurations in the gauge fields implying small fluctuations of the photon fields, $|\theta_{x,\mu}| \ll 2 \pi$.

In the continuum limit~\eq{eq:theta:cont}, the lattice action~\eq{eq:S} of non-singular gauge fields $A_\mu$ becomes the standard Abelian gauge action $S = (1/4 g^2) F^2_{\mu\nu}$ for the photon fields $A_\mu$. To this end, we associate the lattice coupling constant $\beta$ with the coupling constant $g$ in the continuum via the lattice spacing $a$:
\beqn
\beta = \frac{1}{g^2 a}.
\label{eq:beta:a}
\eeqn
Since the lattice coupling~\eq{eq:beta:a} is the dimensionless quantity, the continuum gauge coupling $g$ has the dimension $[g] = {\text{mass}}^{1/2}$ in three spacetime dimensions. The relation~\eq{eq:beta:a} is valid in the weak-coupling regime which also corresponds to large values of the lattice coupling~$\beta$. The weak coupling provides us with a link between the lattice and continuum versions of the model.

\subsection{Monopoles, confinement, and mass gap}

The monopoles in the lattice model~\eq{eq:S} manifest themselves in the form of the strong fields $\theta$ which correspond to large values of the gauge plaquettes $|\theta_P| \sim \pi$. In the continuum limit, such plaquettes lead to singular field-strength tensor $F \sim \theta_P/a^2 \to \infty$. As a result, the continuum action includes singular Dirac lines attached to the Abelian monopoles. A pedagogical introduction to the continuum formulation of compact QED is given in detail in Ref.~\cite{ref:book:Kleinert}.

The monopole charge in the lattice formulation is the gauge-invariant quantity which takes integer values:
\beqn
\rho_x = \frac{1}{2\pi} \sum_{P \in \partial C_{x}} {\bar \theta}_P \, \in \Z\,,
\label{eq:rho:lattice}
\eeqn
where the sum goes over all faces $P$ of an elementary cube $C_x$. The density~\eq{eq:rho:lattice} is expressed via the physical plaquette angle is:
\beqn
{\bar \theta}_P = \theta_P + 2 \pi k_P \in [-\pi,\pi),
\label{eq:bar:theta}
\eeqn
where $k_P \in \Z$ is the integer number.

The world trajectory of a magnetic monopole is a instantaneous since the monopole density is singular in isolated points~\eq{eq:rho:lattice}. Therefore, in two spatial dimensions, the monopole is an instanton-like topological object. The monopoles appear due to the compactness of the gauge group that comes from the invariance of the gauge action~\eq{eq:S} under the discrete transformations of the lattice gauge-field strength: 
\beqn
\theta_P \to \theta_P + 2 \pi n_P, \quad 
\mbox{with} \quad n_P \in \Z.
\eeqn

Thus, the model~\eq{eq:S} describes the dynamics of weak fields of photons and strong fields of monopoles. The photons characterize perturbative fluctuations responsible for a short-distance Coulomb potential between electric charge probes. The monopoles lead to nonperturbative effects such as the long-range linear potential 
\beqn
V(L) = \sigma L\,,
\label{eq:V:L}
\eeqn
%MCH continue
between the oppositely charged probe particles separated by the distance $L$. The linear slope of the potential~\eq{eq:V:L} is given by the tension of a confining string~\cite{Polyakov:1976fu}
\beqn
\sigma = \frac{4 g \sqrt{\varrho}}{\pi}\,.
\label{eq:sigma}
\eeqn
which stretches between the static particle and antiparticle and bounds them into a chargeless particle-antiparticle pair. In Eq.~\eq{eq:sigma}, the string tension is expressed via the mean monopole density~\cite{Polyakov:1976fu},
\beqn
\varrho \equiv \avr{|\rho_x|}_{\mathrm{gas}},
\label{eq:monopole:density}
\eeqn
of a dilute monopole gas. The subscript ``gas'' in the above equation indicates that only the density of the individual (isolated) monopoles is taken into account. The monopoles in rightly bound clusters are ignored in Eq.~\eq{eq:monopole:density}. We will discuss this issue shortly later.

We specially stress the linear behavior of the nonperturbative confining potential~\eq{eq:V:L} because in a monopole–free theory in two spatial dimensions, the potential between electrically charged particle and anti-particle is also formally confining: it is logarithmically rising with the distance. While the logarithmic potential is (weakly) confining, the logarithmic confinement is a trivial result of the reduced dimensionality and it does not reflect any nonperturbative physics.
 
In addition to the linear confinement, the presence of the monopole-antimonopole gas generates the mass gap in the system~\cite{Polyakov:1976fu}:
\beqn
m = \frac{2\pi \sqrt{\varrho}}{g}\,,
\label{eq:mass:gap}
\eeqn
which damps exponentially all correlations at large distances. The photon, for example, becomes massive with the mass given in Eq.~\eq{eq:mass:gap}. 

The string tension~\eq{eq:sigma} and the mass gap~\eq{eq:mass:gap} are derived for the globally neutral Coulomb gas of individual monopoles and antimonopoles. The real gas may contain two type of constituents: (i) isolated monopoles and antimonopoles in the Coulomb component, and (ii) magnetically neutral monopole-antimonopole pairs as well as their clusters. It is the density of the former~\eq{eq:monopole:density} which contributes to the nonperturbative effects, Eqs.~\eq{eq:sigma} and~\eq{eq:mass:gap}, while the density of the monopoles in the neutral pair, expectedly~\cite{Chernodub:2001ws}, does not contribute to the string tension and to the mass gap.

\subsection{Finite-temperature deconfinement}

The compact QED resides in the confining phase at zero temperature. As the temperature of the system raises, two different effects appear: the overall density of monopoles and antimonopoles diminishes while the monopoles and antimonopoles tend to bound into neutral monopole clusters. Both effects contribute to the reduction of the density of free monopoles~\eq{eq:monopole:density}, that diminishes the string tension~\eq{eq:sigma} and the mass gap~\eq{eq:mass:gap}. 

At certain critical temperature  $T = T_c$, all monopoles get bounded so that they exist only in a form of neutral pairs or clusters above $T_c$. As a result, the linear confinement and mass gap generation persist for low temperatures $T < T_c$, while for $T > T_c$ the string tension vanishes and the energy of a pair of static charges behaves logarithmically with their spatial separation. 

In the Wick-rotated theory, the temperature $T$ is associated with the lattice length in the Euclidean time direction $L_t$
\beqn
T = \frac{1}{L_t a}\,,
\label{eq:T:a}
\eeqn
where the lattice spacing $a$ is related to the physical gauge coupling $g$ and lattice gauge coupling $\beta$ via Eq.~\eq{eq:beta:a}.

In addition to the linear slope of the confining potential~\eq{eq:V:L}, the confining properties of the system can be characterized by the Polyakov loop
\beqn
L_{\bs x}(\theta) = \exp \left( i \sum_{x_3 = 0}^{L_t -1} \theta_{x,3}\right),
\label{eq:L:x}
\eeqn
expressed via the time component ($\mu=3$) of the vector gauge field $\theta_{x,\mu} \equiv \theta_\mu(x)$. The sum in Eq.~\eq{eq:L:x} is taken along the Euclidean (imaginary) time direction $\tau \equiv x_3$. The Polyakov loop $L_{\bs x}$ is a spatially local operator defined at a spatial point ${\bs x} = (x_1,x_2)$ and independent of the Euclidean time coordinate $x_3$.

In the thermodynamic limit, the Polyakov loop~\eq{eq:L:x} is an order parameter of the deconfinement phase transition: the vacuum expectation value 
\beqn
\avr{L} =  \frac{1}{L_s^2} \avr{\left|\sum_{x_1 = 0}^{L_s - 1} \sum_{x_2 = 0}^{L_s - 1} L_{x_1,x_2}\right|}\,, 
\label{eq:L:tot}
\eeqn
vanishes in the confinement phase and it takes a nonzero values in the deconfinement phase. Physically, the expectation value of $L_{\bs x}$ is associated with the free energy $F_{\bs x}$ of an isolated static electric charge:
\beqn
e^{- F/T} = \avr{L}\,,
\label{eq:free:energy}
\eeqn
where $T$ is the temperature of the system. According to Eqs.~\eq{eq:beta:a} and \eq{eq:T:a}, the physical temperature $T$, expressed in units of the coupling constant $g^2$,
\beqn
\frac{T}{g^2} = \frac{\beta}{L_t}\,,
\label{eq:T:phys}
\eeqn
is a linear function of the lattice gauge coupling $\beta$. In the low-temperature confinement phase, the order parameter $\avr{L_{\bs x}}$ is zero, implying that the free energy of a separate charge~\eq{eq:free:energy} is infinite, so that an isolated electric charge cannot exist. In the high-temperature  deconfinement phase, the order parameter and the associated free energy do not vanish implying the existence of free electric charges (deconfinement).

The (de)confining properties of compact U(1) gauge theory may be contrasted with the features of non-Abelian (Yang-Mills) gauge theories in 3+1 dimensions. Both these theories possess a similar phase structure consisting of a linearly confining low-temperature phase and a deconfined phase at a finite temperature. The deconfining phase transition of a Yang-Mills theory is associated with a spontaneous breakdown of a global $\Z_N$ center symmetry of the underlying SU(N) gauge group. In the pure SU(N) gauge theory, the transition is of the second order for two colors and is of the first order for the number of colors three or greater. 

On the contrary, the phase transition in the compact U(1) gauge theory in 2+1 dimensions is not associated with a center group as the U(1) symmetry remains unbroken in both phases. Moreover, the transition has an infinite order that maintains all local observables analytical as the system passes the critical temperature. The deconfinement is associated with binding of (anti)monopoles into magnetically neutral compact pairs and clusters at high temperature: the Coulomb gas of magnetic monopoles becomes a gas of neutral magnetic dipoles at $T=T_c$.
This type of critical behavior is known as the Berezinskii--Kosterlitz--Thouless (BKT) transition~\cite{Berezinskii:1971fst,Berezinskii:1971scn,Kosterlitz:1973xp}.

The BKT transition is associated with a loss of the confinement property at high temperature because the weak fields of the neutral magnetic dipoles cannot lead to a disorder of the Polyakov loop. At low temperature, the disorder is driven by the long-range fields of the individual magnetic monopoles and anti-monopoles. 

On the practical level, the deconfinement temperature at a given lattice may be calculated as the position of the peak of the susceptibility 
\beqn
\chi_L = \avr{L^2} - \avr{L}^2
\label{eq:chi:L}
\eeqn
of the order parameter~\eq{eq:L:x} (as for example, performed in Ref.~\cite{Chernodub:2001ws} and many others). Alternatively, one may determine the pseudocritical temperature via location of the maximal slope of the order parameter $\avr{L}$ itself. The critical temperature is given by the thermodynamic limit of either of the pseudocritical temperatures calculated at finite spatial volumes.

Notice that we are always working with finite volume lattices, therefore it is more suitable to call these quantities as pseudocritical, while we will use the word ``critical'' for shortness.

At the level of the topological defects and associated the BKT-type restructuring of the monopole ensembles, the determination of the critical temperature is much less clear. Although this question may be eventually resolved via a thorough determination of the neutral monopole clusters and appropriate correlations~\cite{Chernodub:2001ws}, the visual difference between a gas of individual monopoles and antimonopoles at the low-temperature side of the transition and loosely-bound magnetic dipoles at the high-temperature size of the transition remains quite obscure.  

This paper aims to identify the phase transition temperature using the machine learning techniques concentrated only on the dynamics of the monopoles. In our approach, the neural network treats the monopole ensembles as three-dimensional images (holograms) and tries to identify the deconfining phase transition as a point where the monopole gas becomes a magnetic-dipole gas. 

\begin{table*}[!htb]
    \centering
    \begin{tabular}{c|cccccc}
         lattice
            & $\beta_c$
            & $\beta_c^{\mathrm{raw}}$
            & $\delta\beta$
            & $a$
            & $b$
            & $\nu$
        \\
\hline        \\
         $4 \times 16^2$
            & $1.811(2)$
            & $1.801(2)$
            & $0.186(4)$
            & $0.317(1)$
            & $0.153(4)$
            & $0.485(22)$
        \\
         $6 \times 16^2$
            & $1.977(4)$
            & $1.960(4)$
            & $0.250(6)$
            & $0.256(3)$
            & $0.127(5)$
            & $0.557(42)$
        \\
         $8 \times 16^2$
            & $2.105(7)$
            & $2.078(7)$
            & $0.323(6)$
            & $0.216(3)$
            & $0.110(5)$
            & $0.560(49)$
        \\
         $4 \times 32^2$
            & $1.931(3)$
            & $1.924(3)$
            & $0.173(5)$
            & $0.276(2)$
            & $0.144(5)$
            & $0.501(36)$
        \\
         $6 \times 32^2$
            & $2.142(6)$
            & $2.131(6)$
            & $0.247(5)$
            & $0.233(3)$
            & $0.135(6)$
            & $0.384(50)$
        \\
         $8 \times 32^2$
            & $2.285(11)$
            & $2.270(10)$
            & $0.306(8)$
            & $0.193(5)$
            & $0.113(7)$
            & $0.382(80)$
    \end{tabular}
    \caption{\raggedright The best-fit parameters from fitting of the expectation value from the Polyakov loop~\eq{eq:L:fit} as well as the value of the (pseudo)critical coupling constant $\beta_c$ obtained from Monte-Carlo simulations.}
    \label{tab:MC-betac}
\end{table*}

\subsection{Details of numerical simulations}

We work with cubic Euclidean lattices $L_t \times L_s^2$ subjected to periodic boundary conditions along all three directions. In our simulations, we take various asymmetric configurations with $L_t = 4,6,8$ and $L_s = 16,32$. 

The configurations of the gauge field are generated with the help of a Hybrid Monte Carlo algorithm. We use standard Monte-Carlo methods improved by molecular dynamics algorithms~\cite{ref:Gattringer} which include a second-order minimum norm integrator~\cite{ref:Omelyan}. Long autocorrelation lengths in Markov chains are eliminated following Ref.~\cite{ref:Gattringer}. We apply a self-tuning adaptive algorithm in order to control the acceptance rate of the Hybrid Monte-Carlo in a reasonable range between 0.70 and 0.85.

We generated $1.1\times 10^6$ trajectories for each value of the coupling constant $\beta$. The thermalization has been performed for $10^5$ trajectories (200 configurations), after which we used 2000 configurations for measurements separated by 500th trajectories. It is more than enough for eliminate correlation between configuration.

\begin{figure}[ht]
	\centering
	\includegraphics[scale=0.325]{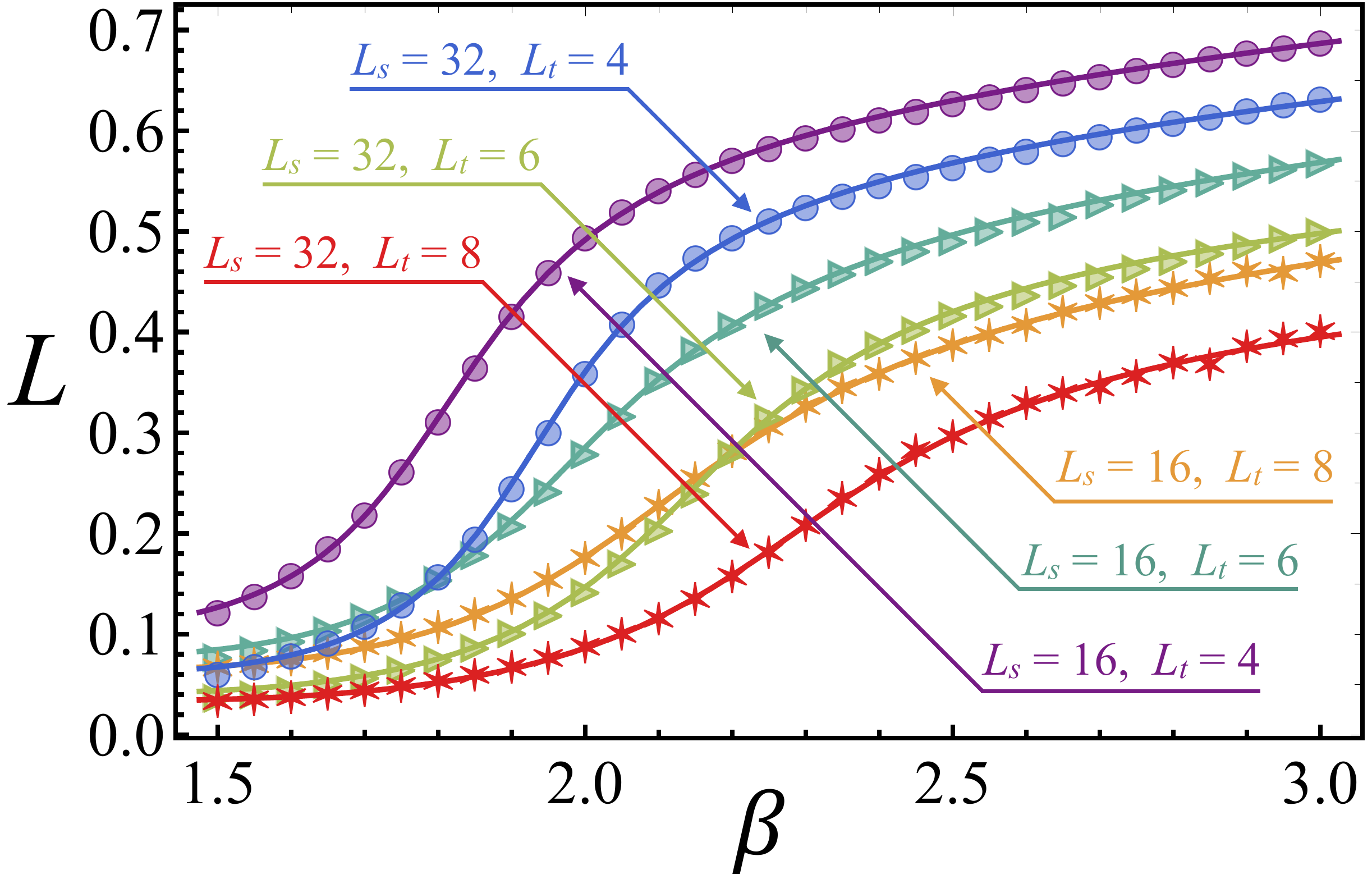}
	\caption{\raggedright The best fits of the expectation value of the Polyakov loop by the function~\eq{eq:L:fit}.}
	\label{fig:fits}
\end{figure}

We calculate numerically the vacuum expectation value~\eq{eq:L:tot} of the Polyakov loop~\eq{eq:L:x} at a dense set of lattice gauge couplings $\beta$ and fit the result by the following function:
\beqn
L^{\mathrm{fit}}(\beta) = a + b \, \beta^\nu \arctan \biggl(\frac{\beta - \beta_c^{\mathrm{raw}}}{\delta \beta}\biggr),
\label{eq:L:fit}
\eeqn
where $a$, $b$, $\beta^{\mathrm{raw}}$ and $\delta \beta$ are the fitting parameters. The pseudocritical value of the coupling $\beta_c$ is then computed as the maximum of the best-fit function~\eq{eq:L:fit}. The quantity $\delta \beta$ characterizes the width of the pseudocritical deconfining transition. The best fits are shown in Fig.~\ref{fig:fits} while the corresponding best fit parameters along with the corresponding values of $\beta_c$ are shown in Table~\ref{tab:MC-betac}.

\section{Neural network }
\label{sec:neural}

We are aiming to build a neural network with the purpose to discriminate monopole configurations in the confining and deconfining phase, and to determine the phase transition point. The machine should learn how to distinguish between the two phases of the theory looking only at the configurations that encodes the positions and charges of the topological defects. The monopole configurations of the compact gauge theory are produced by the Monte-Carlo algorithm which simulates the physical properties of the model from first principles. We would like the neural network to learn the monopole properties, to understand them, and to make predictions based on the knowledge acquired during the learning phase. 

More specifically, the objective of our work is to train the neural network at a part of the configurations and to make predictions using new configurations which were not seen by the neural network during the training phase. In order to make the training and prediction phases as much independent as possible, we train the network on the configurations of the lattice size $L_t = 4$ and $L_s = 16$ and make the predictions at a different set of sizes $L_t = 4, 6, 8$ and $L_s = 16, 32$. Then, from the predicted Polyakov loop $ L $ and phase $\phi$ (to be defined later), we derive the critical coupling $\beta_c$ of the phase transition and compare it with the value given by the first-principle Monte Carlo simulation. In this way the neural network is trained to see the difference between the confining and deconfining phases of the compact electrodynamics.

Traditional neural networks are made of three types of objects: layers of neurons (also called units or filters depending on the layer type), connections (which strengths are called \textit{weights}), and activation functions.
The layers are usually arranged sequentially, with each pair of adjacent layers linked by connections.
A neuron is a real number whose value is determined by a linear combination of the neurons from the previous layer, described by the weights of the connection (and the type of the layer), to which the activation function is applied.
In the simple case of fully connected layers, each layer can be represented as a vector and the connections between two layers by a matrix, such that each layer is given by the result of the activation function applied on each component of the vector obtained from the matrix multiplication of the weights by the previous layer.
The first and last layers correspond respectively to the inputs and outputs (targets).

\textit{Training} a neural network consists of tuning the weights until good results are obtained.
The simplest approach is called supervised learning, where a gradient descent is performed in order to minimize the differences between the predicted values (last layer) and the expected values.
These differences are measured according to a distance, or \textit{loss} function, appropriate for the problem at hand.
The architecture of a neural network is determined by the layers (types, number of neurons…), by the choice of activation functions and by all parameters needed to define the network; it is kept fixed during the training.
Moreover, the use of a gradient descent implies that all quantities appearing in the expression of the loss must be differentiable. 

In general, it is hard to guess directly the best architecture: for this reason, different architectures are considered successively in a process known as \textit{hyperparameter tuning}, which alternates changing the structure and training the network.
At the end, the performances of all the different architectures are compared to find the best.
This phase is also used to find the best training parameters (including the gradient descent algorithm).
The reason for splitting this procedure in two steps is the following: minimizing a loss function using a gradient descent requires the parameters to be continuous variables.
This is the case of the connection weights, but not of many other parameters defining the networks (such as the number of neurons per layer).
Another reason is that it is easy to find how the weights enter in the expression of the predictions, such that one can easily take the gradient; this is not the case of the other parameters which do not appear directly in the expressions (for example, the number of neurons or the form of the 
activation function).

\subsection{Configurations and targets}

Technically, the monopole configuration is represented as three-color hologram encoded as a $3d$ tensor of size $(L_t, L_s, L_s)$. The entries of the hologram are $+ 1$, $-1$, or $0$ corresponding to a monopole, antimonopole, or an empty space, respectively. One may imagine it as a $3d$ image with one channel taking three possible values, for example, black, white and gray. Since we want to work at different lattice sizes, we need to make sure that the network can take holograms of arbitrary sizes as input.

The goal is to extract the critical temperature of the phase transition from the predictions. Therefore, we focus on the most relevant quantities for this purpose: the absolute value of the Polyakov loop $ L $ (order parameter) defined in \eqref{eq:L:tot} (we omit the symbol $\avr{\cdot}$) and the phase label:
\begin{equation}
	 \phi =
		\begin{cases}
			0 & \text{confined},
			\\
			1 & \text{deconfined}.
		\end{cases}
\end{equation}

For continuous quantities such as the Polyakov loop $ L $, the prediction can be taken directly to be the neuron value in the layer with a trivial activation function is not needed since it would just changed the value of the weights before.
However, the neural network is trained not to predict the phase label, but rather the probability $p(\phi)$ to find $\phi = 1$. Indeed, the gradient descent requires that each activation function is differentiable: getting a value $\phi = 0$ or $\phi = 1$ can be achieved by using the Heaviside step function, which is not a differentiable function.
Instead, the sigmoid function
\begin{equation}
    \sigma(x)
        = \frac{1}{1 + e^{- x}}
\end{equation}
is differentiable and produces an output between $0$ and $1$ which is interpreted as a probability.
Moreover, this choice offers some flexibility: for example, one can tune the probability cut-off to favor one label (to counter-balance a bias) or to spot uncertainty on the classification (at the phase transition).
% 1505.06279

It is straightforward to add several outputs to a neural network. This step generically improves the performance of the network by enforcing the stability and generalization: indeed, adding additional layers forces the layers at the beginning of the network to look for more universal features.
We added several secondary variables that could be leveraged by the neural network (even if this helped only marginally in our case): the real and imaginary parts of the Polyakov loop, the temporal $U_t$, spatial $U_s$ and average $U$ plaquettes, the temperature $\beta$ and the monopole density $\rho$.

\begin{figure*}[!htb]
	\centering
	\includegraphics[scale=0.51]{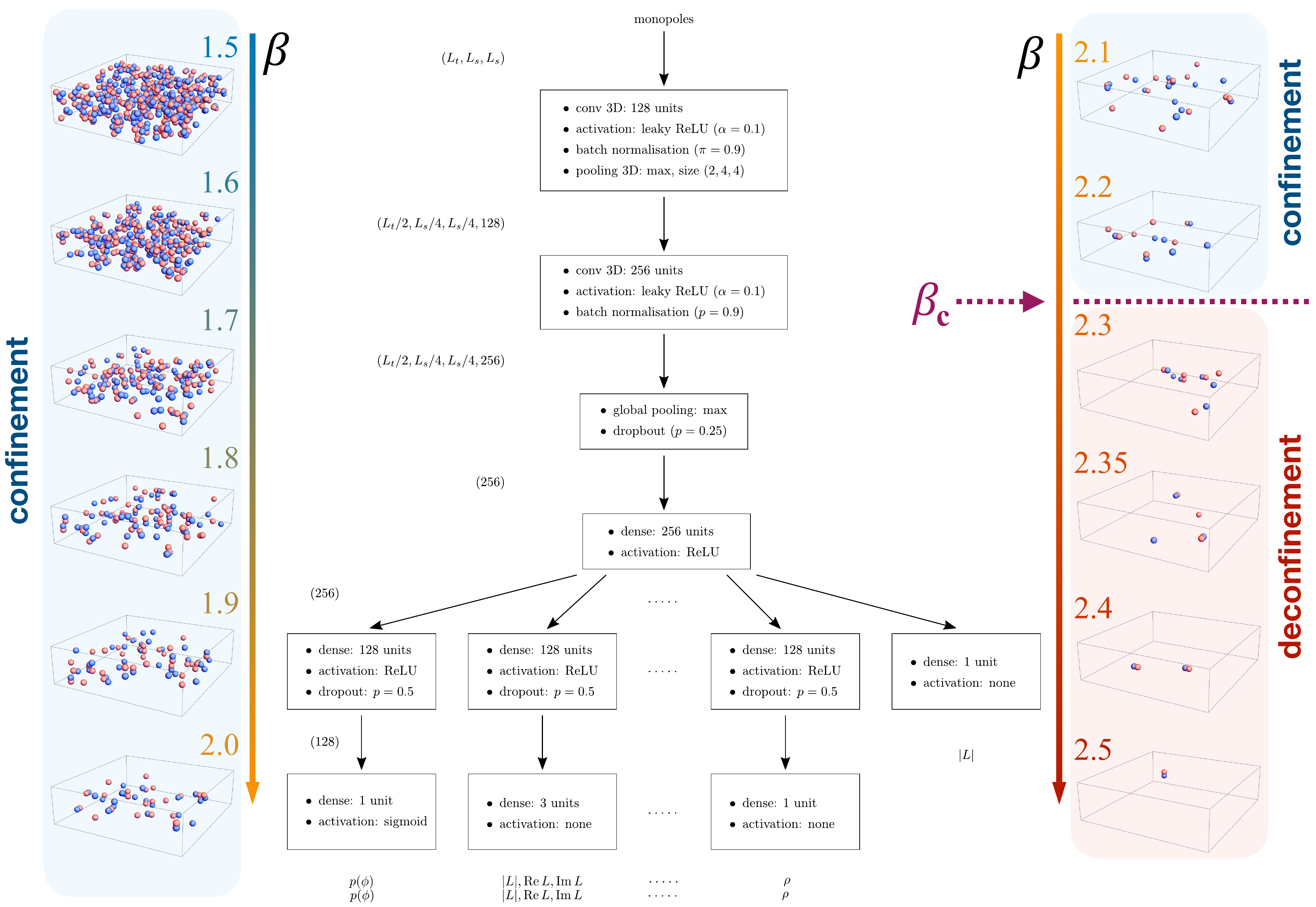}
	\caption{\raggedright Neural network model and the typical examples of the monopole configurations as the function of the increasing temperature $\beta$ or, according to Eq.~\eq{eq:T:phys}, rising temperature $T$.}
	\label{fig:nn-model}
\end{figure*}

\subsection{Structure}

We can now describe the internal structure of the network (Fig.~\ref{fig:nn-model}). Most ingredients are standard and we refer the reader to the literature~\cite{ref:review:1, ref:book-ml:1, ref:book-ml:2, ref:book-ml:3} for more details.

Since the input is a $3d$ image, the first two layers are $3d$ convolutional layers with $128$ and $256$ filters of size $(2, 4, 4)$ in order to account for the translational symmetry of the lattice.
The effect of convolutions is to give the holograms as many channels as the number of filters.
The first layer is followed by a $3d$ max pooling\footnote{This procedure can be understood as a coarse graining operation, where a block of $(2, 4, 4)$ neurons is replaced by a single neuron whose value is the maximal one of the block. Other pooling operations (such as averaging) are possible and give similar results.} with size $(2, 4, 4)$; this procedure reduces the image size which makes the training faster (information is preserved in the channels generated by the convolution).
The second layer is followed by a global max pooling, which keeps only the maximal value of the image for each channel.
This is necessary in order to pass the data to fully connected dense layers: while convolutional layers can take holograms of arbitrary sizes as input, this is not the case of the dense layers.
Ultimately, this ensures that the network can be fed with holograms of any size (i.e.\ the monopole configurations can live on different lattices).
After the global pooling operation comes a dense layer with $256$ units and leaky ReLu activation (slope $\alpha = 0.1$).
At this stage, the network branches in five directions, one for each group of variables we want to compute: $p(\phi)$, $( L , \Re L, \Im L)$, $(U, U_s, U_t)$, $\rho$, $\beta$.
Each branch contains a dense layer with $128$ units and ReLu activation, which is followed by the final dense layers which output the predictions (thus, these layers have $1$ or $3$ units depending, no activation except a sigmoid for the probability prediction).
The idea behind this structure is to share the layers until some point to encourage learning more general (and hopefully robust) features, while the final layers can specialize in computing its output.
Since early layers can tend to forget or learn useless information for deep networks, we added an auxiliary output of $ L $ before the branching: the corresponding loss is added with a smaller weight of $0.3$ (since the network is expected to be less accurate in the early layers, we should penalize it less).
When computing all possible quantities, the network has roughly $1.3$M parameters.

Standard techniques have been used to improve further the convergence (both in terms of speed and performance).
Batch normalization (with momentum $0.9$) has been added after the convolutional layers.
Dropout layers have been added after the last convolutional layer (keep probability $0.25$) and after the last dense layer of each branch (keep probability $0.5$): this randomly deactivates some links during the training, forcing the network to not rely on specific neurons, but to find more generic properties and to achieve some redundancy.

The different outputs have different scales: this can disturb the network as targets with higher absolute values would contribute more to the loss, implying that the network will put more weight on getting them correctly while ignoring the other targets.
For this reason, it is useful to standardize all targets by subtracting the mean and diving by the standard deviation.
Note that the mean and standard deviation are computed from the training set only.
A second motivation is the intuition that this could make the network less sensitive to changes in the lattice size.

\subsection{Training}

The loss of the network measures its performances by comparing the predicted values to the real ones (the latter are computed by Monte Carlo). The loss is given by the sum of the mean squared errors of continuous variables plus the cross-entropy for the phase (binary) classification.
During the training phase, a weight regularization loss is added: it is proportional to the $L_2$-norm of the weight of the neural network.
This procedure helps us to reduce the numbers of parameters, reducing the risk of over-fitting and improving generalization.
The neural network is then trained by performing a gradient descent in order to minimize the loss.
Hence, the network learns to reproduce the expected values as outputs while having as small weights as possible.
In order to put more incentive in getting the correct absolute Polyakov loop $ L $ and the phase probability $p(\phi)$, we can weight the different terms of the loss function to penalize less for incorrect values of the secondary variables.
However, this did not give results sensibly different, to we took a weight of $1.$ for all quantities.
% The weights are: $1.$ for $p(\phi)$ and $ L $ and $1.$ for the other quantities.

One should be careful when comparing losses: 1) during training, the losses include the $L_2$-term, which is removed when evaluating the model after training; 2) the losses are proportional to the number of parameters of the model, and thus it depends on the precise structure of the network.
% As a consequence, losses during and after training generically differ because the regularization term is not included and dropout is removed.

The network is trained with early stopping: we monitor the performance on a validation dataset (not used for training) and we stop training when the performance does not improve anymore, rolling back to the best network.
This is another form of regularization since the network has less time to adapt to the training dataset.
The maximum number of epochs is set to $75$.
We used a batch size of $256$ and the Adam optimization method~\cite{ref:book-ml:1}.
The training set is randomly shuffled before each epoch.

The neural network output for the phase can be interpreted as a probability: and it is necessary to define how to extract the phase from it.
We use the following decision function:
\begin{equation}
	\label{eq:dec-fn}
    \phi =
    	\begin{cases}
    		0, \qquad & p(\phi) < p_c, \\
    		1, & p(\phi) \ge p_c,
    	\end{cases}
\end{equation}
where $p_c$ is the probability threshold.
The standard choice would be $p_c = 0.5$, but $p_c$ must be interpreted as an hyperparameter, on the same footing as the other hyperparameters of the network.
In particular, it can be used to fight the bias towards the size $4 \times 16^2$ and we have found that the value $p_c = 0.85$ leads to good values of the critical temperature.
Changing $p_c$ amounts to find a compromise between precision and recall (see Table~\ref{tab:classif-metrics} which will also be described in the analysis below).

The training is done for the lattice $4 \times 16^2$ with two datasets:
\begin{itemize}
	\item $2000$ configurations for each lattice coupling in the range $\beta \in [1.5, 3]$ with the step $\Delta \beta = 0.05$ (i.e., $31$ values in total)

	\item $100$ configurations for each $\beta \in [0.1, 2.2]$ with the step $\Delta \beta = 0.1$ ($22$ values)
\end{itemize}
We stress that there is a single training phase for the complete training set including configurations at all beta: since samples are random shuffled before each epoch, batches contain different combinations of configurations at different beta.
Validation is performed for the same lattice with a third dataset: $200$ configurations for each $\beta \in [1.5, 2.5]$ with $\Delta \beta = 0.05$ ($21$ values).
The predictions are evaluated for all lattices $6 \times 32^2$, $8 \times 32^2$, $4 \times 32^2$, $6 \times 32^2$, $8 \times 32^2$, using for each a dataset having $200$ configurations for each $\beta \in [1.5, 2.5]$ with $\Delta \beta = 0.05$ (like the validation set). Results for the validation set are also given.

\section{Results: Neural network\\ learns monopoles and confinement}
\label{sec:results}

The aim of this section it to describe the results how the neural network may learn the dynamics of monopoles and predict the various observables as well as the position of the deconfinement transition. 

Primarily, we are interested in the order parameters of the (de)confinement phase transition. There are two such parameters: the Polyakov loop $ L $ which comes from the field theory and the phase label $\phi$ which is a ML-analogue of the order parameter. Alternatively, we may also consider the susceptibility Polyakov loop~$\chi_L$ given in Eq.~\eq{eq:chi:L} at the field-theoretical side and the probability $p(\phi)$ at the side of the neural network. We provide the predictions of the neural network for the critical temperature $\beta_c$, the monopole density $\rho$, and,  for completeness, the mean plaquette $U$.

\subsection{General observations}

To start with we notice that the training stage takes circa $6$ minutes (on a GPU Nvidia GeForce RTX2080 Ti), while all predictions are obtained in few seconds.

\begin{figure}[ht]
	\centering
	\includegraphics[scale=0.35]{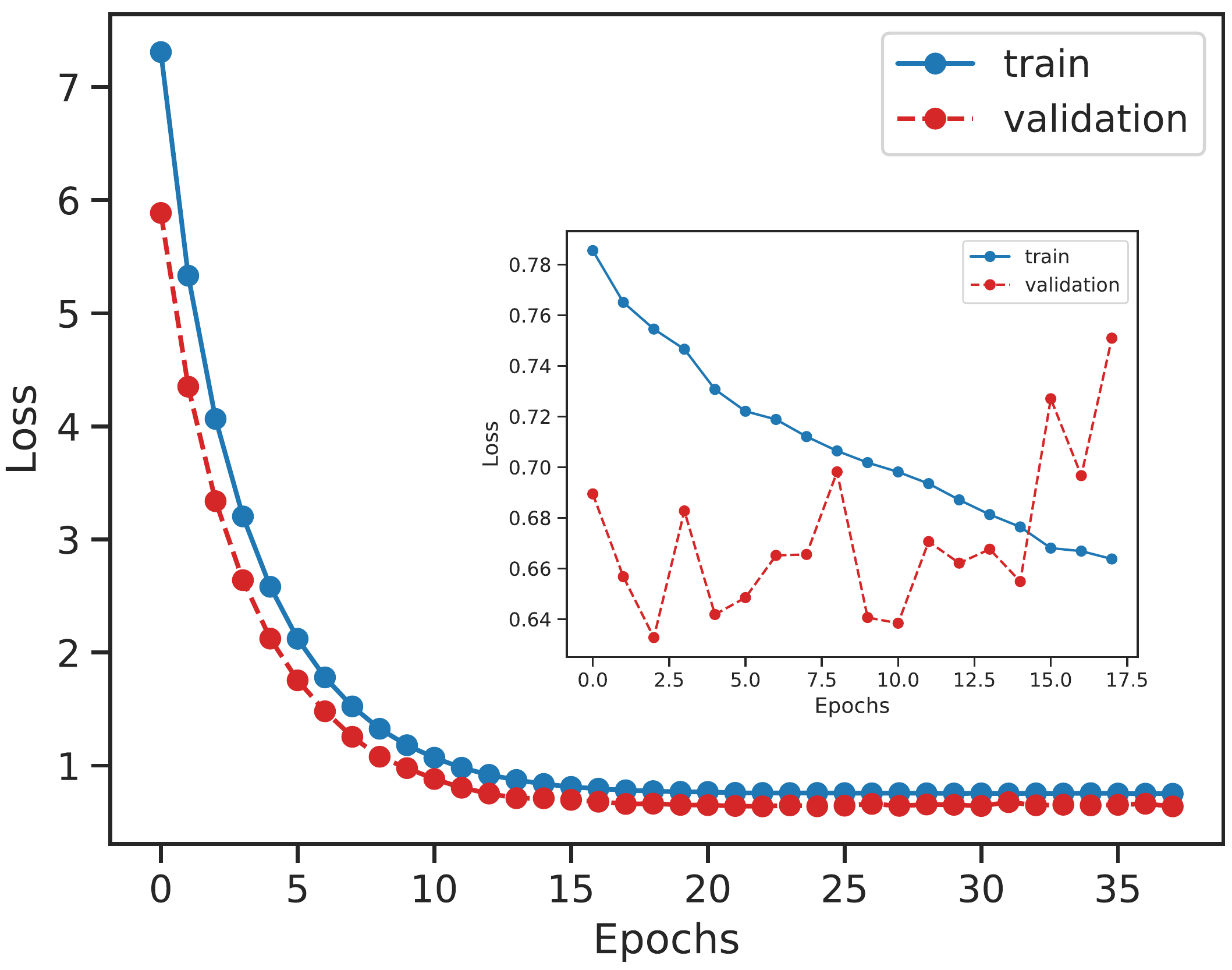}
	\caption{\raggedright Training curve (evolution of losses for the training and validation sets with time). The inset: the training curve without regularization.}
	\label{fig:training-curve}
\end{figure}

Convergence to the best model is achieved after a dozen of epochs as it is illustrated in Figure~\ref{fig:training-curve}. In practice, we find that the performance is quite stable under changes of parameters (for example, the choice of optimizer, the use of the scaling or not using it, the choice of layer sizes, etc).

\begin{figure}[ht]
	\centering
	\includegraphics[scale=0.35]{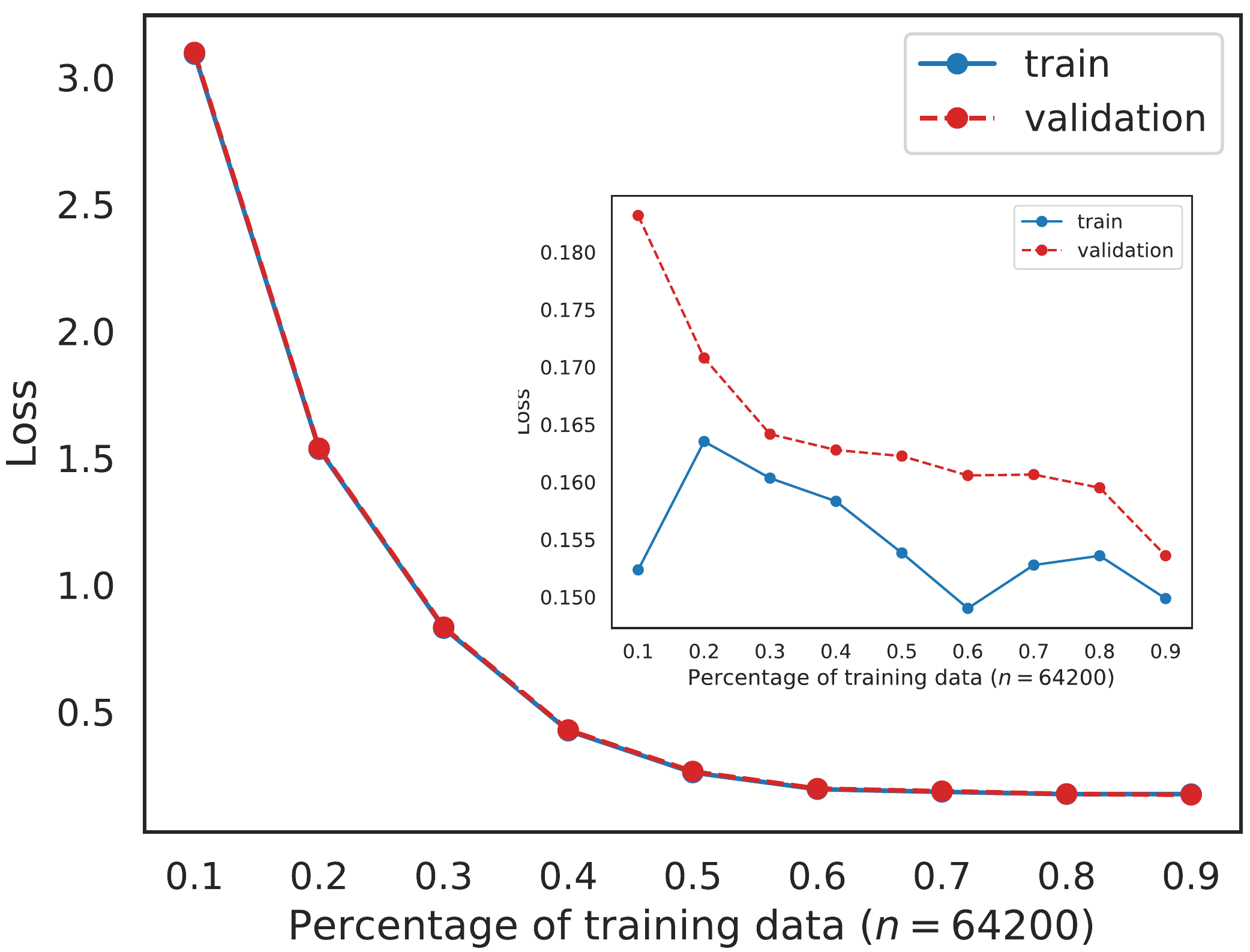}
	\caption{\raggedright Learning curve (evolution of losses for the training and validation sets of different sizes). The inset: the learning curve without regularization.}
	\label{fig:learning-curve}
\end{figure}

Interestingly, we find that the validation and training losses reported on the training curve (loss evolution during the training, Figure~\ref{fig:training-curve}) and on the learning curve (loss evolution by changing the ratio of training/validation data, Figure~\ref{fig:learning-curve}) are residing on top of each other. Such agreement between the two curves is rather uncommon,\footnote{Figure~\ref{fig:learning-curve} represents a type of an ideal curve predicted by the theory for an optimally regularized network. In practice, it almost never happens to encounter such a curve.} especially at early stage of training. This property indicates the absence of over-fitting and under-fitting, implying that the neural network architecture is very well adapted for the task.
The reason for such a nice agreement comes from an efficient regularization, as it can be seen by comparing with the same network without regularization (no $L_2$-term and no dropout, with the result shown in the insets of Figures~\ref{fig:training-curve} and~\ref{fig:learning-curve}, respectively).
The flattening of the learning curve (Figure~\ref{fig:learning-curve}) for high ratio indicates that adding more training data is unlikely to improve the performances.

\begin{figure}[ht]
	\centering
	\subcaptionbox{$\phi$ ($p_c = 0.5$)}{\includegraphics[scale=0.25]{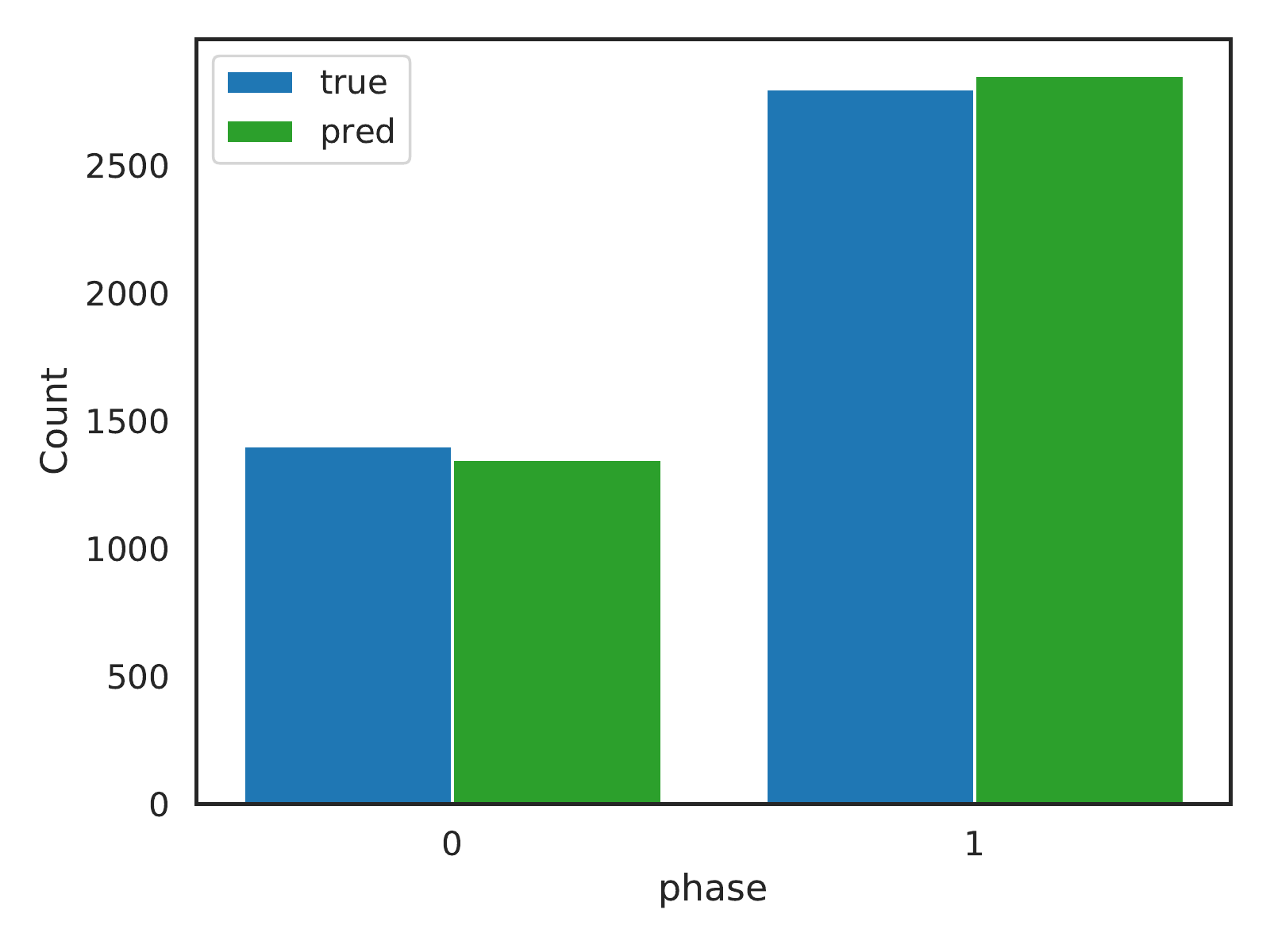}}
	\subcaptionbox{$\phi$ ($p_c = 0.85$)}{\includegraphics[scale=0.25]{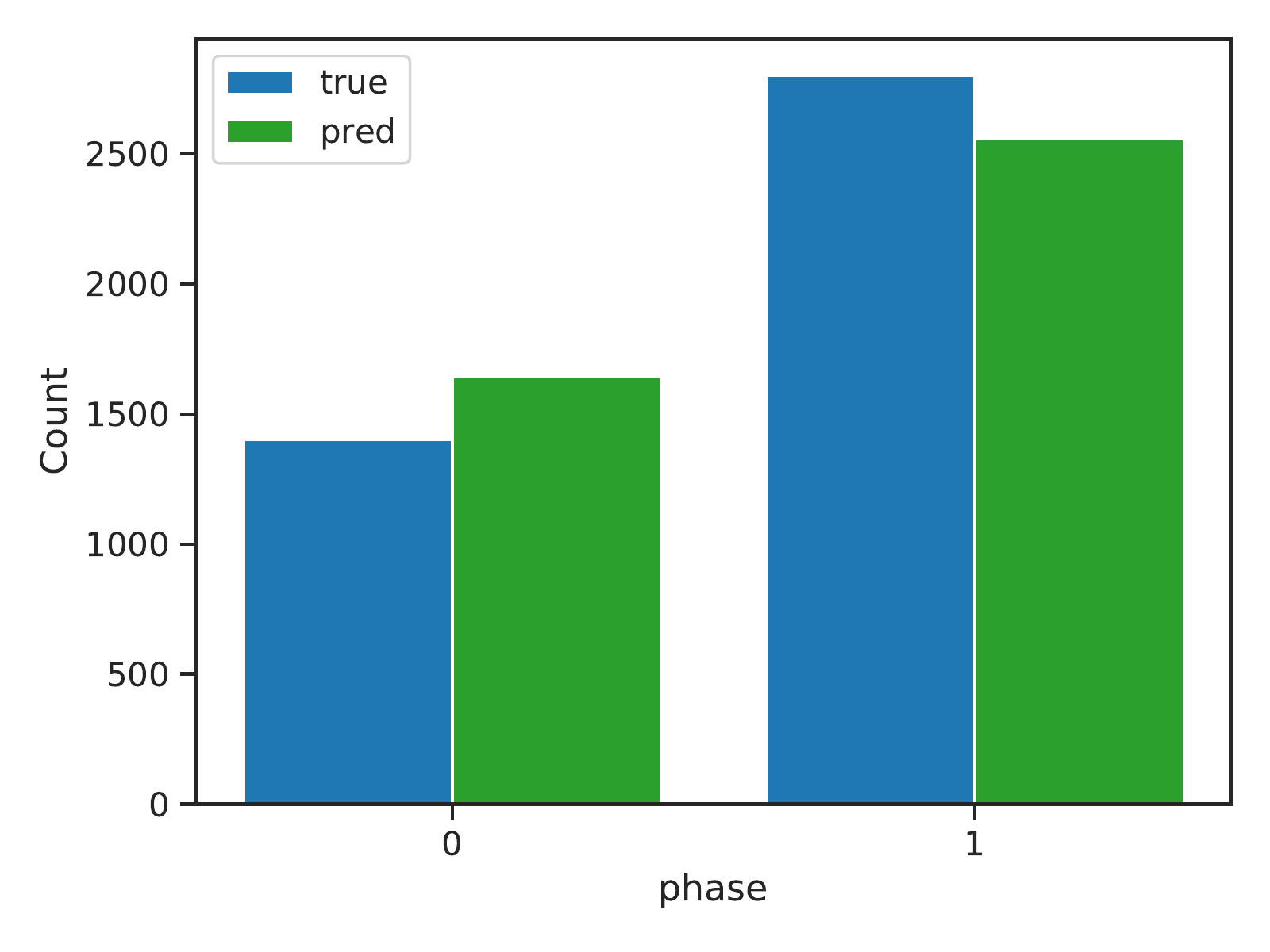}}
	\subcaptionbox{$ L $}{\includegraphics[scale=0.25]{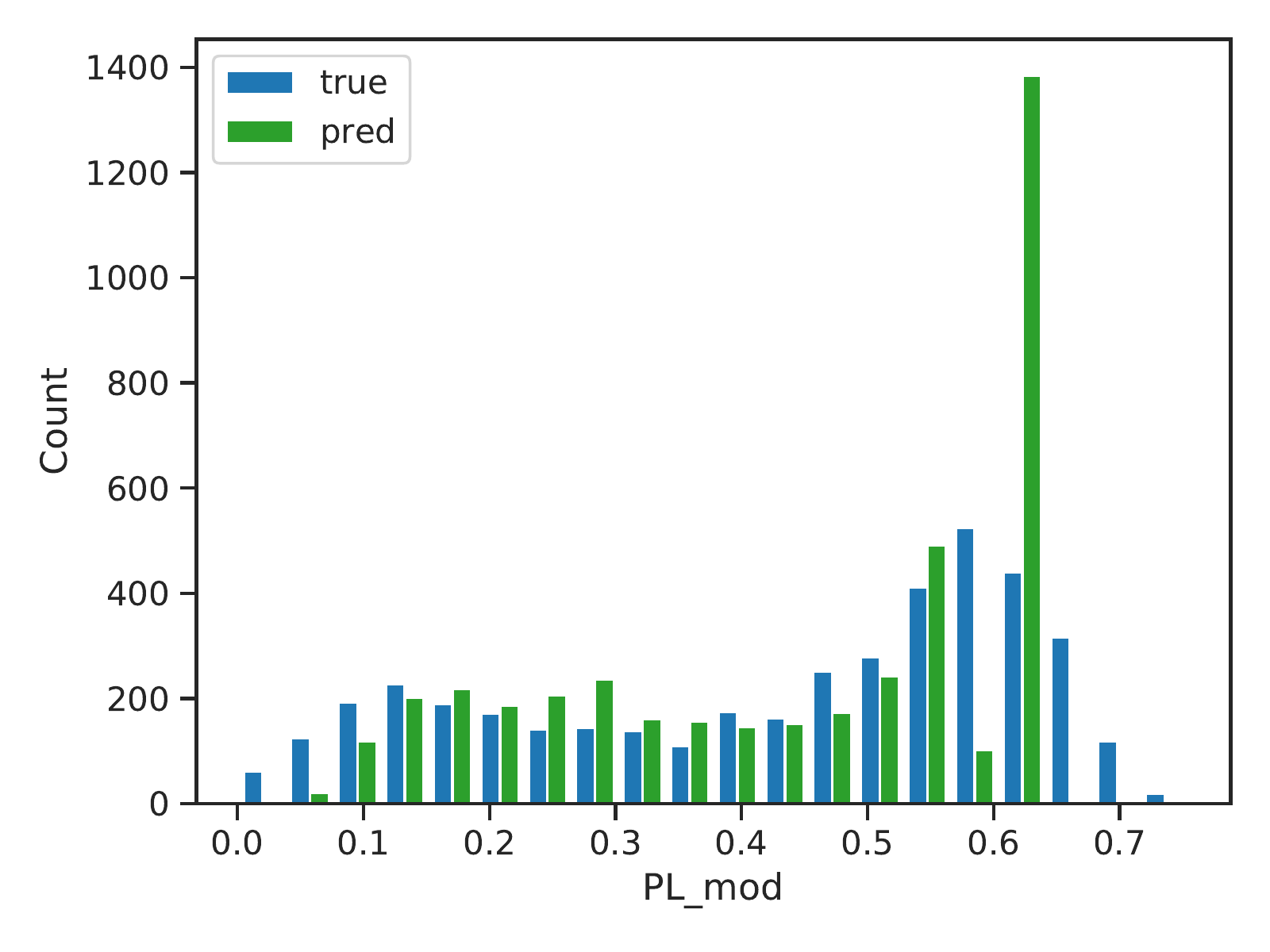}}
	\subcaptionbox{$\rho$}{\includegraphics[scale=0.25]{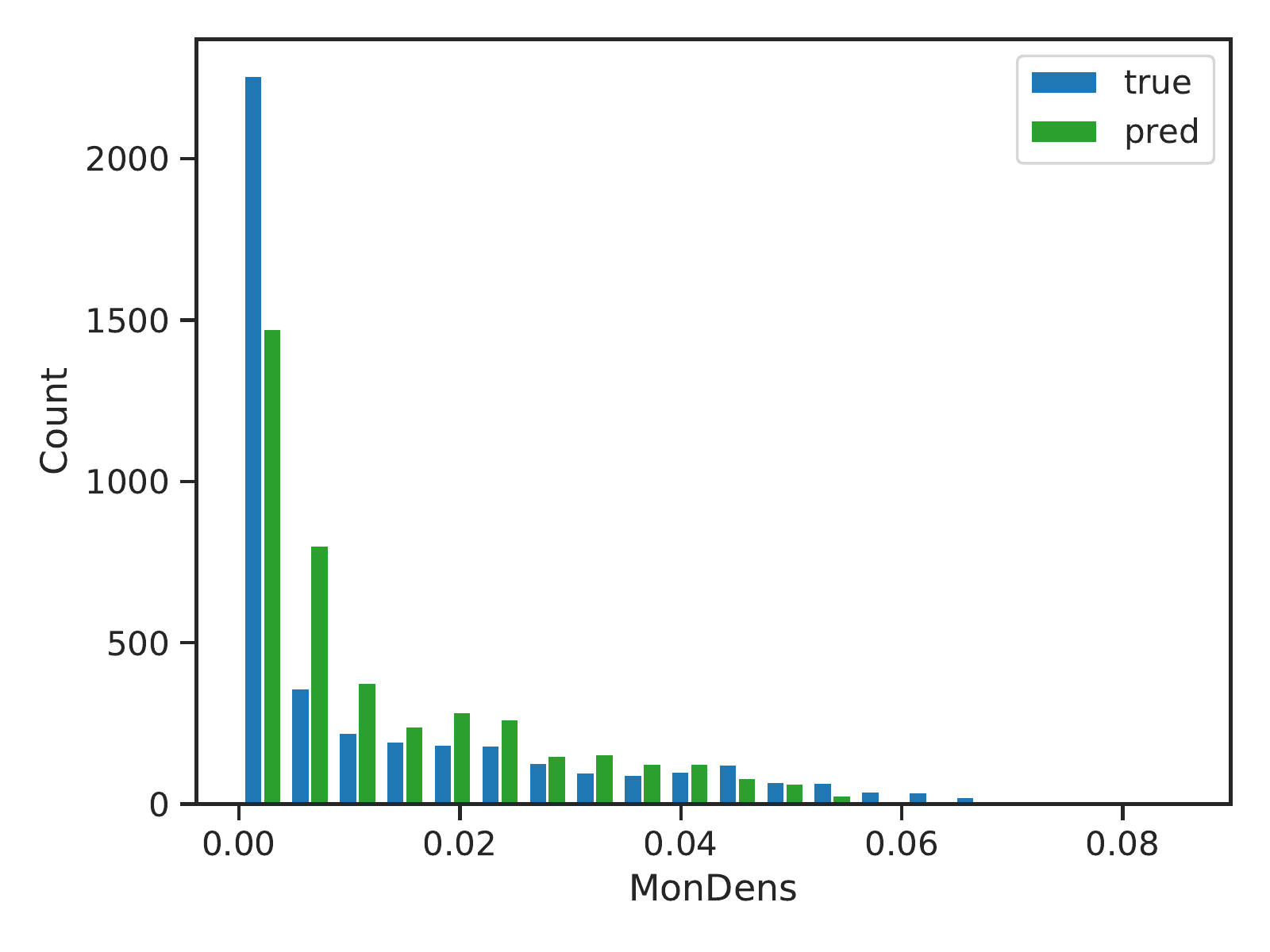}}
	\subcaptionbox{$\beta$}{\includegraphics[scale=0.25]{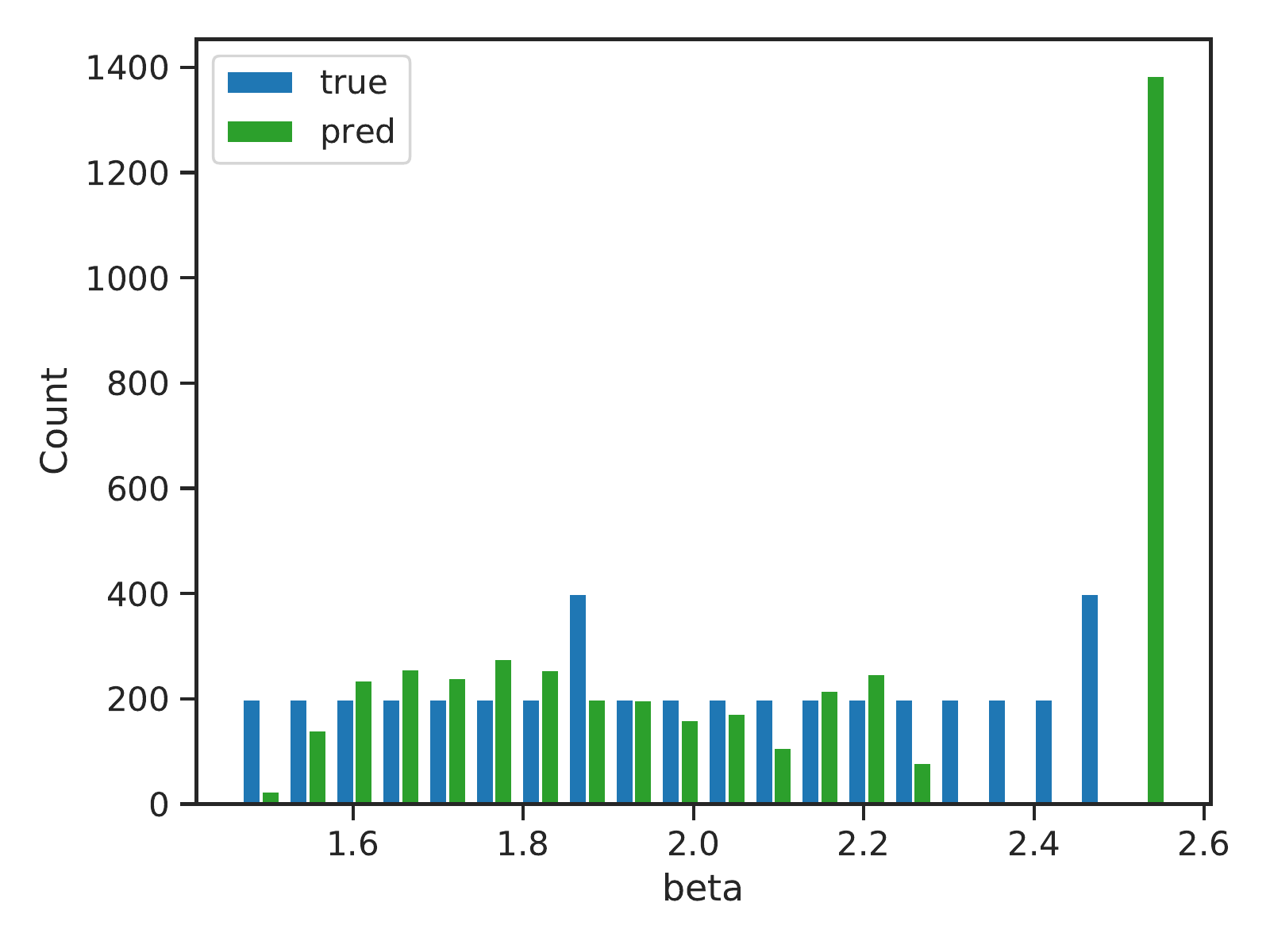}}
	\subcaptionbox{$U$}{\includegraphics[scale=0.25]{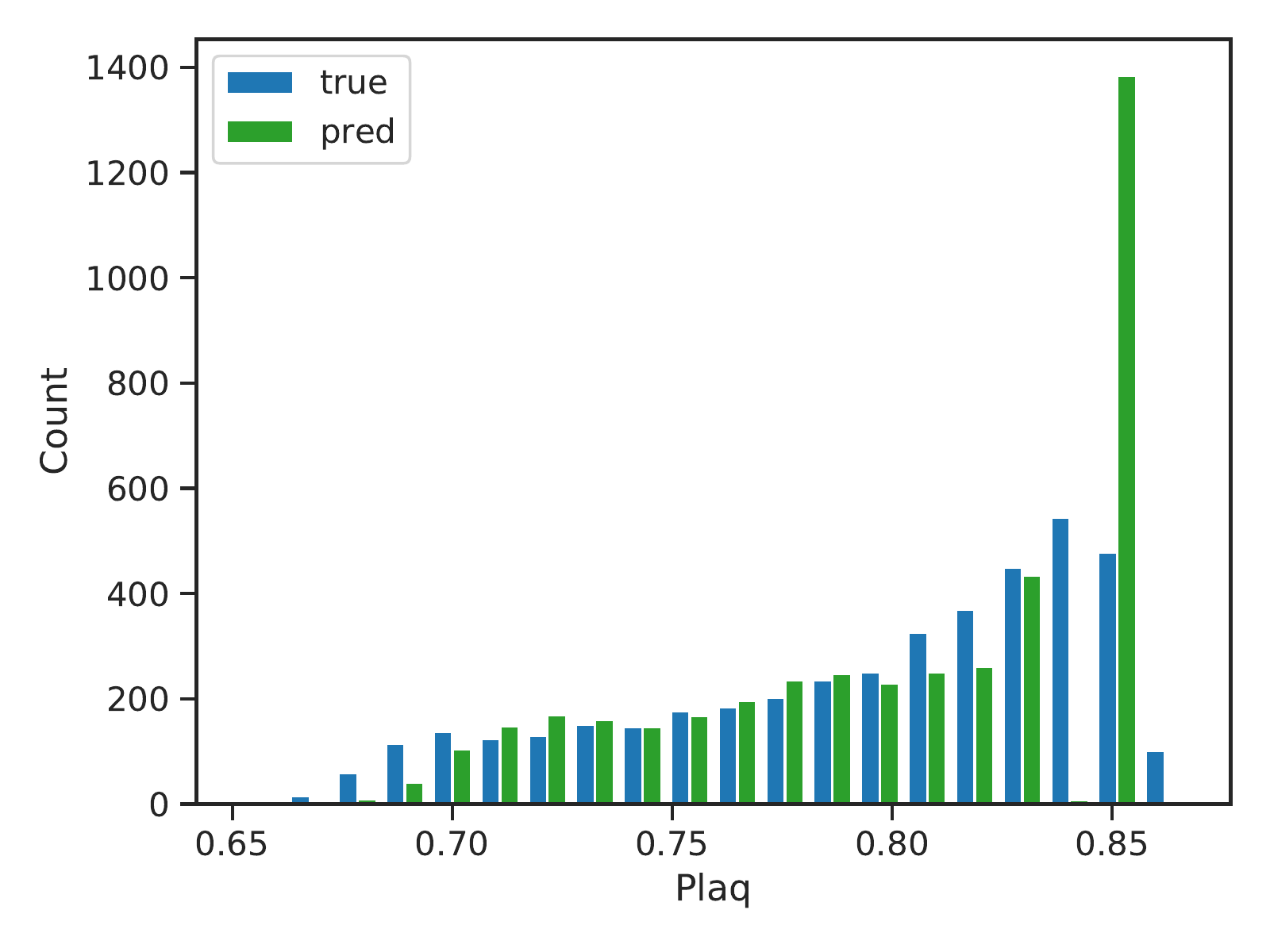}}
	\caption{\raggedright Comparison of the MC and ML distributions of various quantities.}
	\label{fig:mc-ml-distrib}
\end{figure}

\begin{figure}[!htb]
	\centering
	\subcaptionbox{$\beta$}{\includegraphics[scale=0.25]{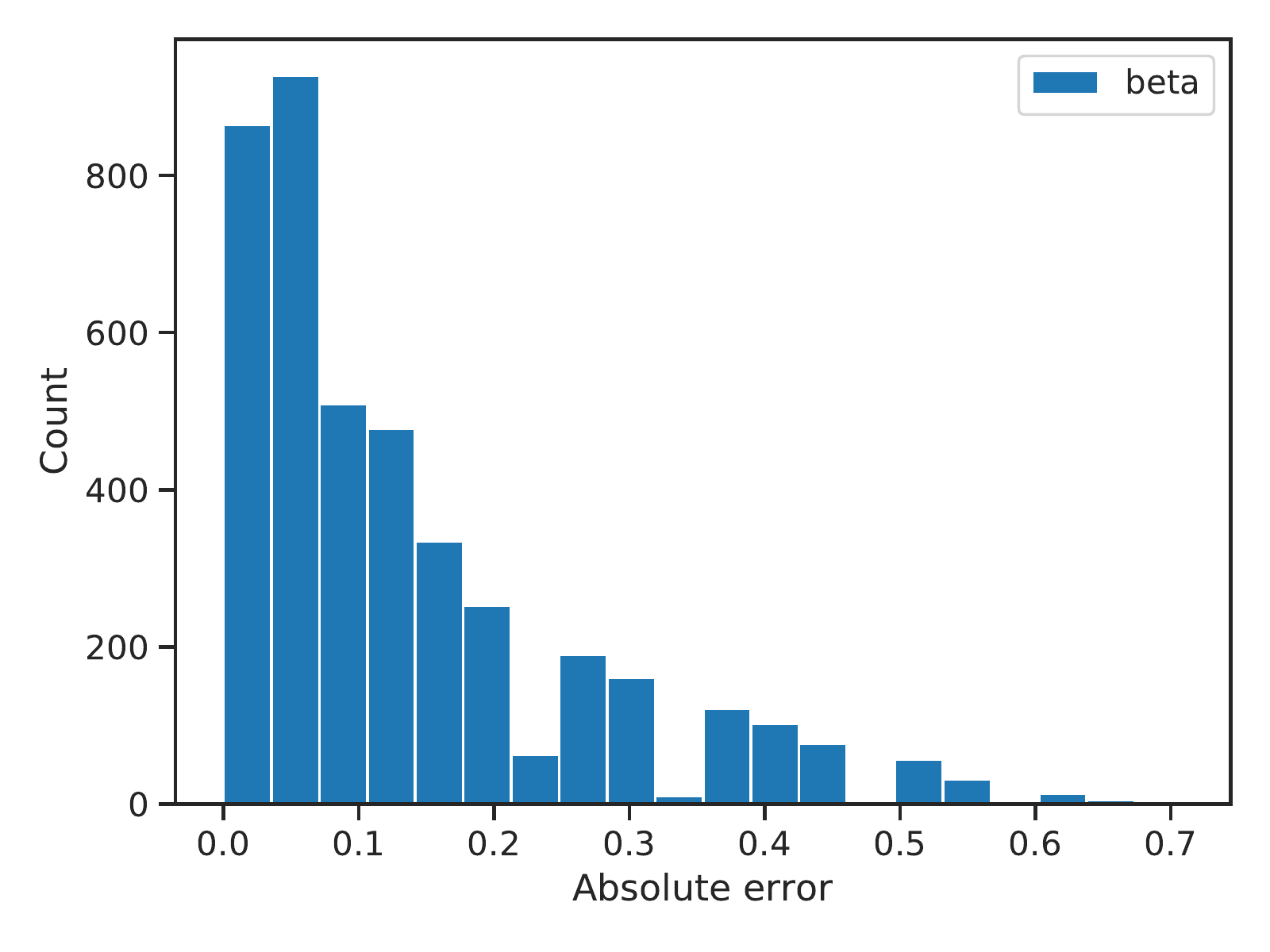}}
	\subcaptionbox{$\rho$}{\includegraphics[scale=0.25]{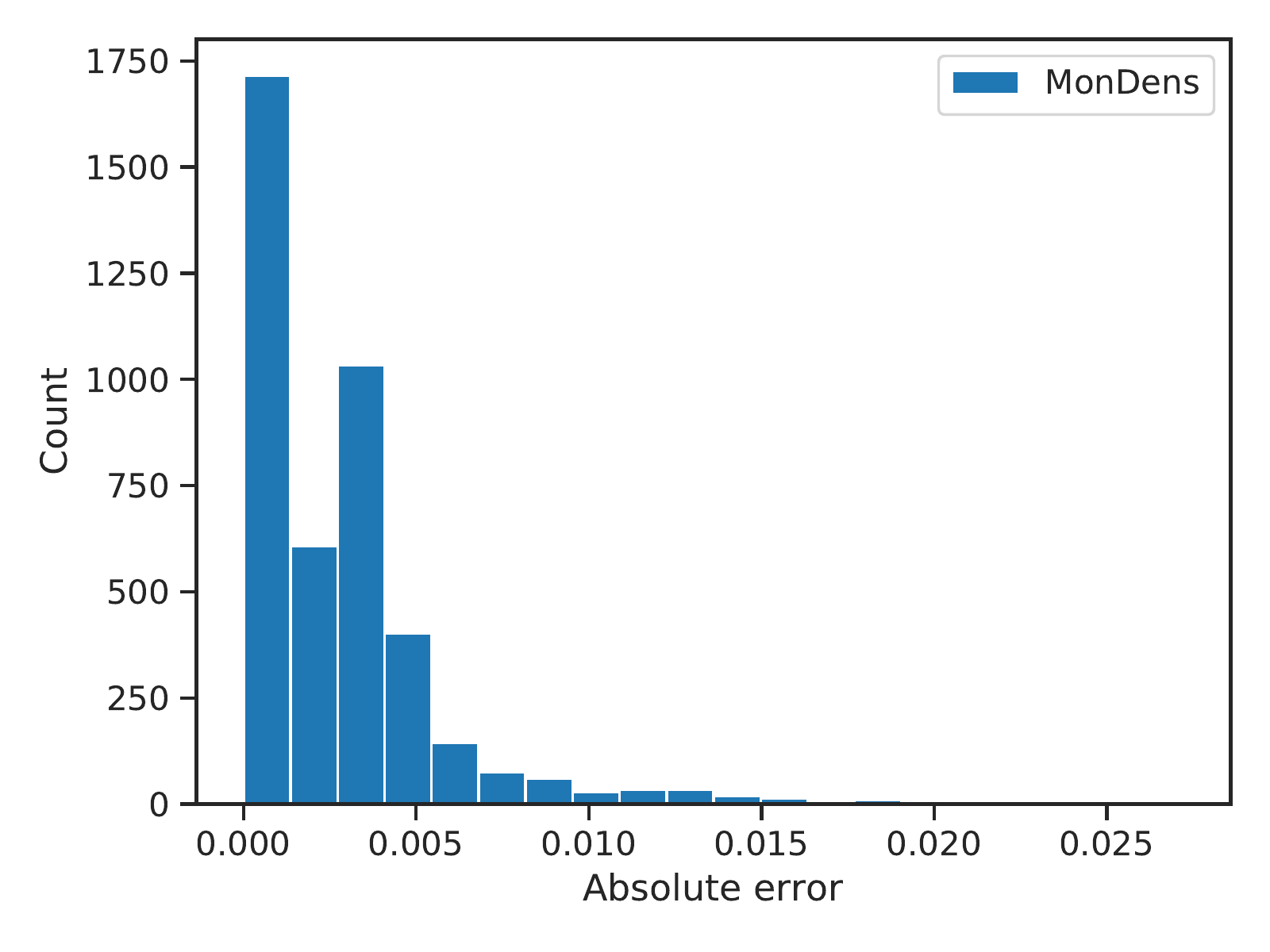}}
	\subcaptionbox{$\phi$ ($p_c = 0.5$)}{\includegraphics[scale=0.25]{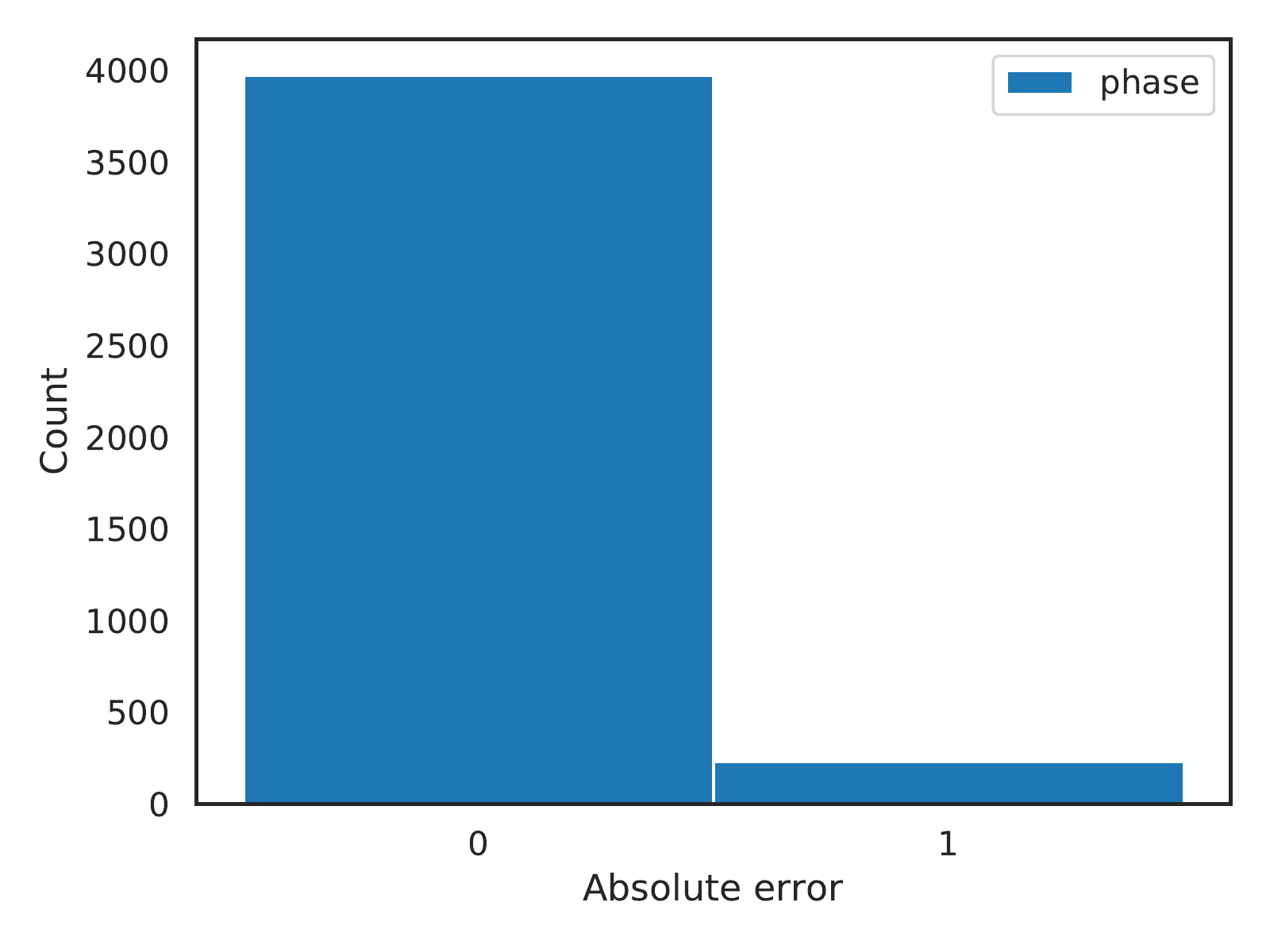}}
	\subcaptionbox{$\phi$ ($p_c = 0.85$)}{\includegraphics[scale=0.25]{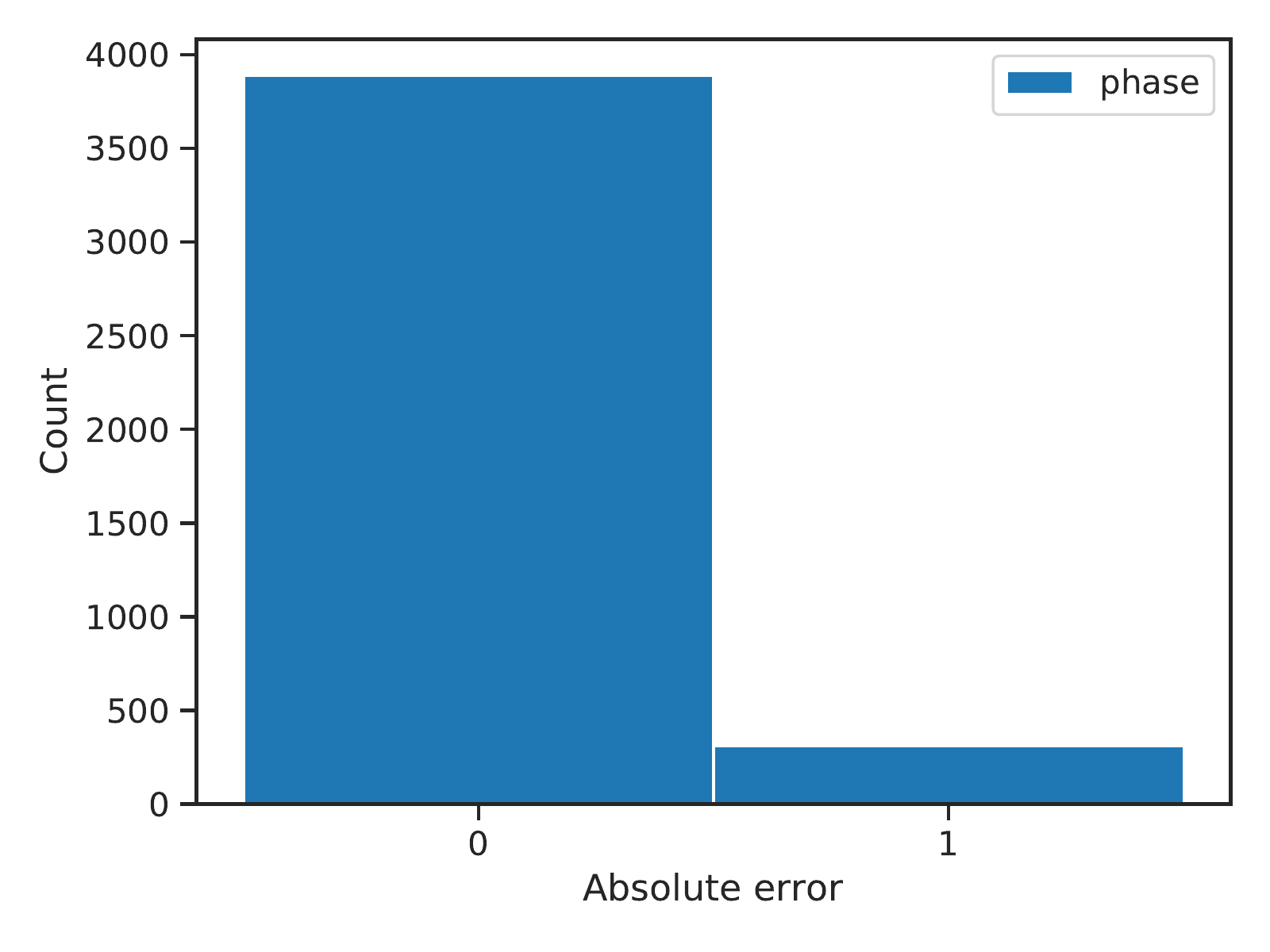}}
	\subcaptionbox{$U$}{\includegraphics[scale=0.25]{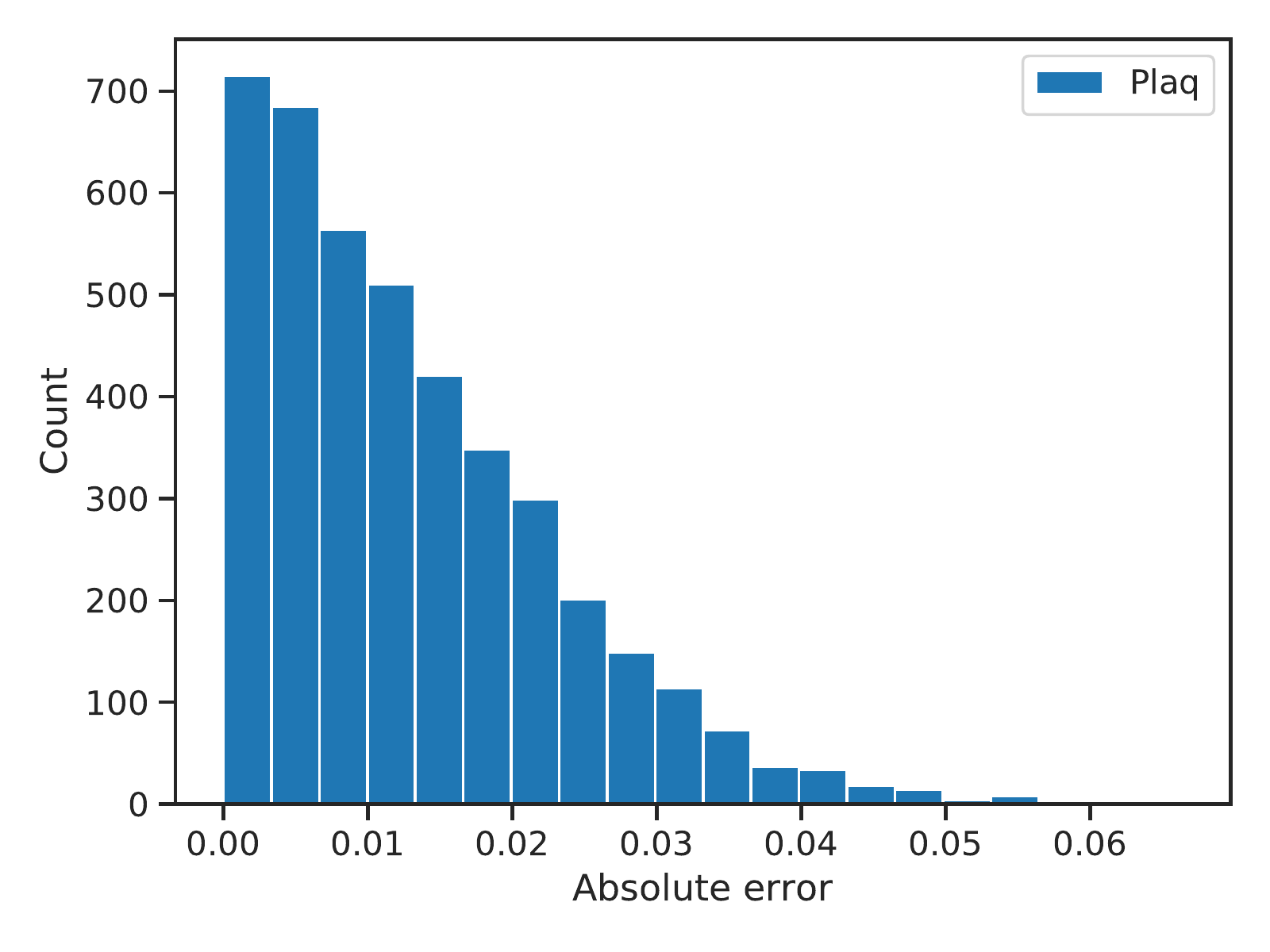}}
	\subcaptionbox{$ L $}{\includegraphics[scale=0.25]{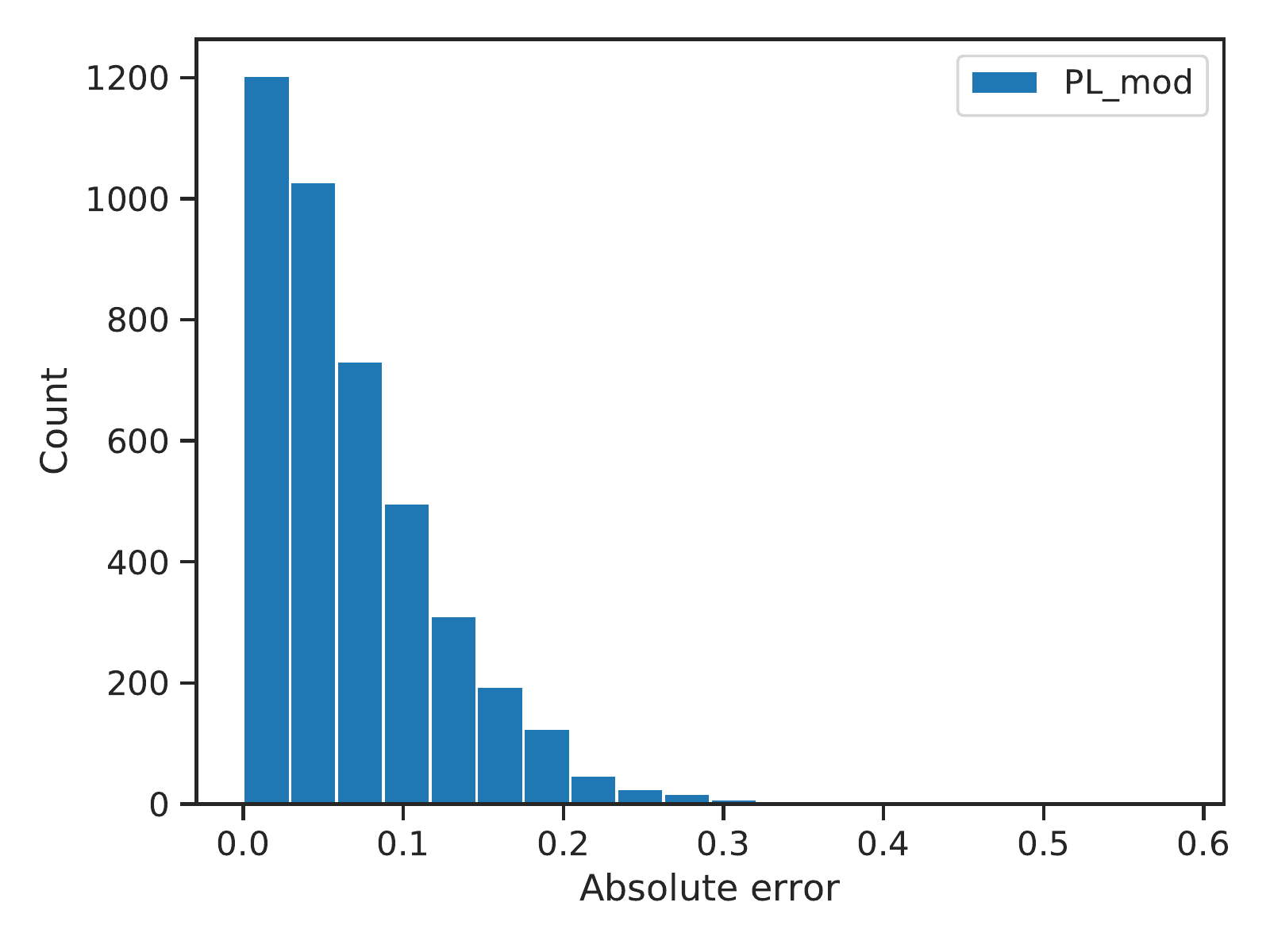}}
	\caption{\raggedright Comparison of absolute errors on the predicted distributions.}
	\label{fig:mc-ml-errors}
\end{figure}

The comparison between the Monte Carlo and ML distributions of the phase label $\phi$, the mean values of the Polyakov loop $ L $, the monopole density $\rho$ and the average plaquette $U$ are given in Figure~\ref{fig:mc-ml-distrib}. We also give the same comparison for the predicted value of the coupling constant $\beta$: the neural network reads a configuration of the magnetic monopoles and predicts the value of the coupling constant $\beta$ that should correspond to this particular configuration.\footnote{Note that despite we challenge the neural network to compute $\beta$, we always use the real $\beta$ when studying the temperature dependence of the predicted quantities.}
The corresponding errors are given in Figure~\ref{fig:mc-ml-errors}. For continuous quantities, the prediction accuracy is summarized in Table~\ref{tab:rmse} in the form of the root mean square error (RMSE). For the phase label $\phi = 0,1$, we characterize the performance metric in Table~\ref{tab:classif-metrics} in terms of the quantities
\beqs
\beqn
 \mathrm{accuracy} & = & 
 \frac{\mathrm{TP} + \mathrm{TN}}{\mathrm{All}}, \\
 \mathrm{precision} & = & 
 \frac{\mathrm{TP}}{\mathrm{TP} + \mathrm{FP}},  \\
 \mathrm{recall} & = & \frac{\mathrm{TP}}{\mathrm{TP} + \mathrm{FN}},
\eeqn
 \label{eq:metrics}
\eeqs
where ``All'' corresponds to the total number of cases, of which ``TP'' means the number of true positive, ``TN'' is true negative, ``FP'' is false positive, ``FP'' is false positive. 

\begin{figure*}[!htb]
	\centering
	\subcaptionbox{MC}{\includegraphics[scale=0.35]{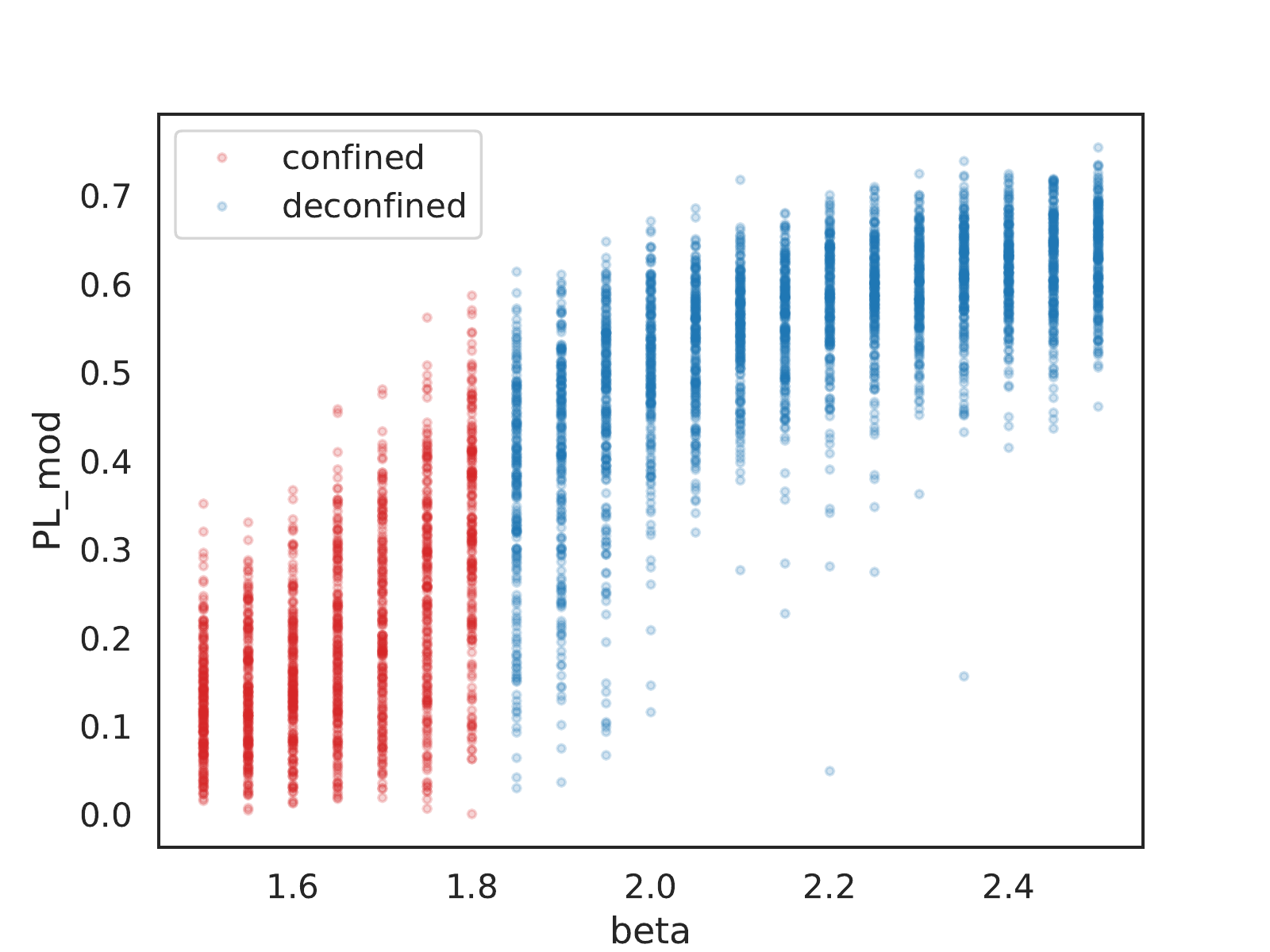}}
	\subcaptionbox{ML, $p_c = 0.5$}{\includegraphics[scale=0.35]{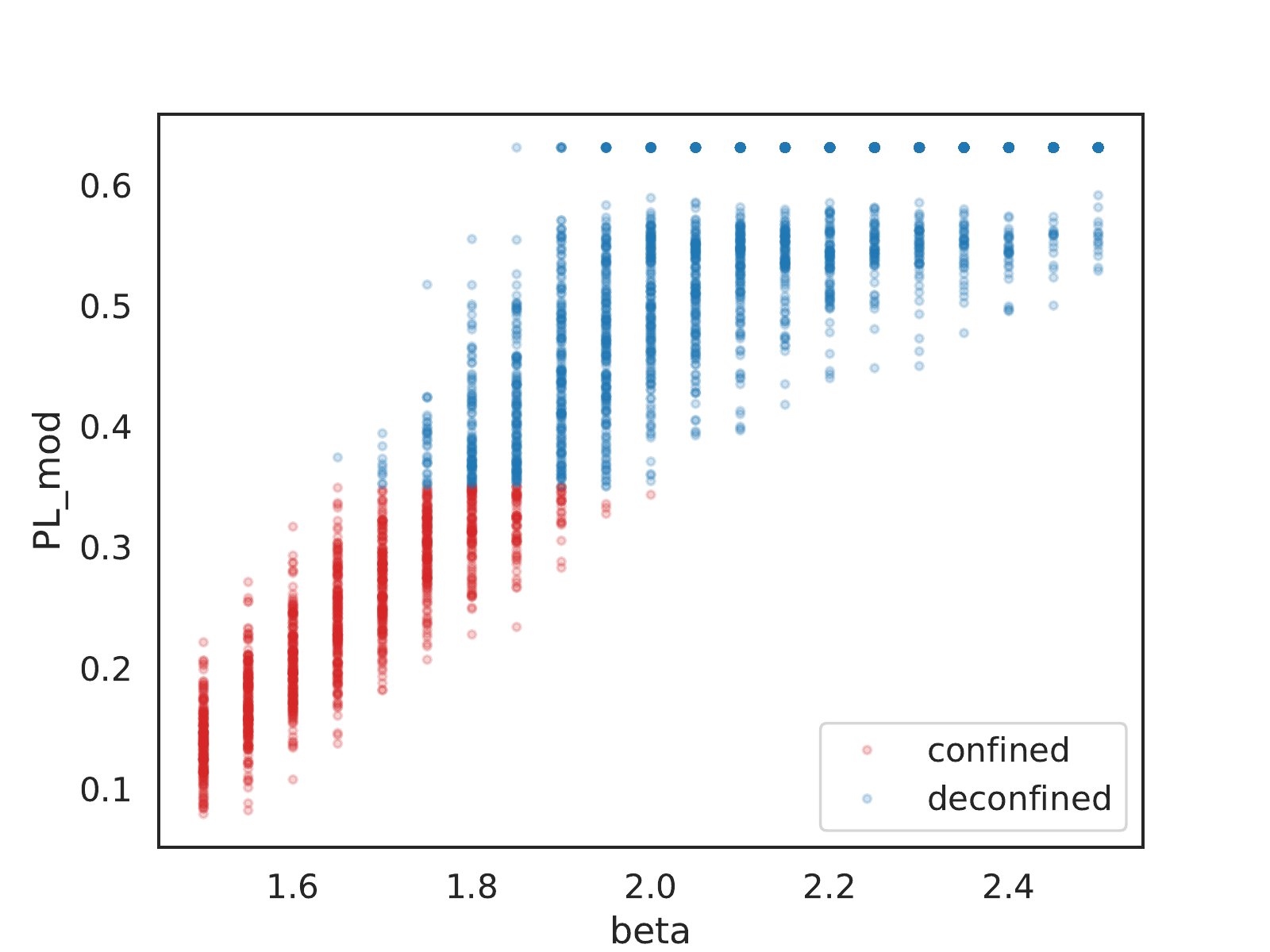}}
	\subcaptionbox{ML, $p_c = 0.85$}{\includegraphics[scale=0.35]{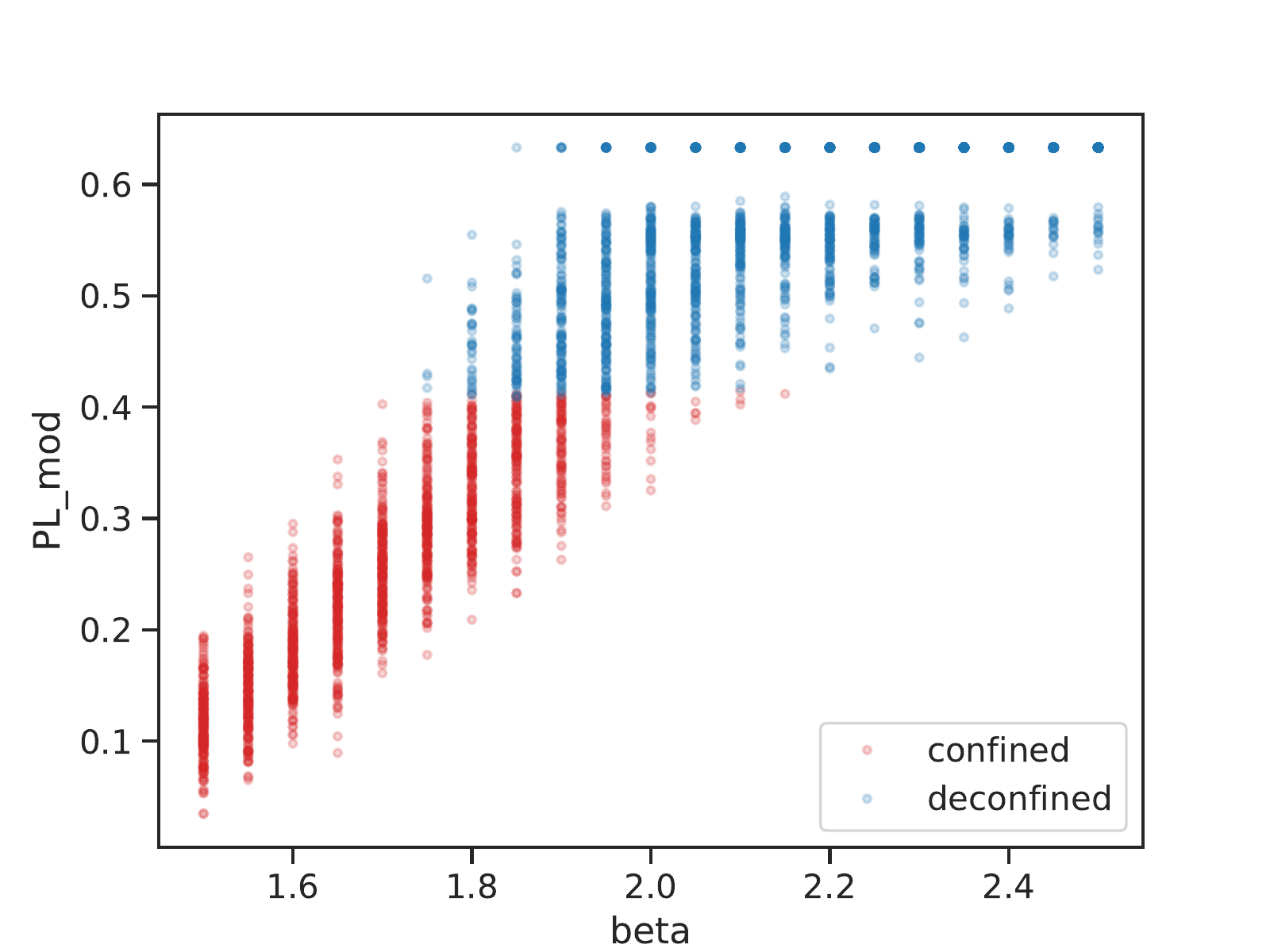}}
	\caption{\raggedright The Polyakov loop $ L $ in terms of $\beta$ for each configuration for $L_t = 4$ and $L_s = 16$, with the phase $\phi$ (red: $p = 0$, confined, blue: $p = 1$, deconfined).}
	\label{fig:temp-L-phi-4-16}
\end{figure*}

\begin{figure}[ht]
	\centering
	\subcaptionbox{$4 \times 16^2$}{\includegraphics[scale=0.25]{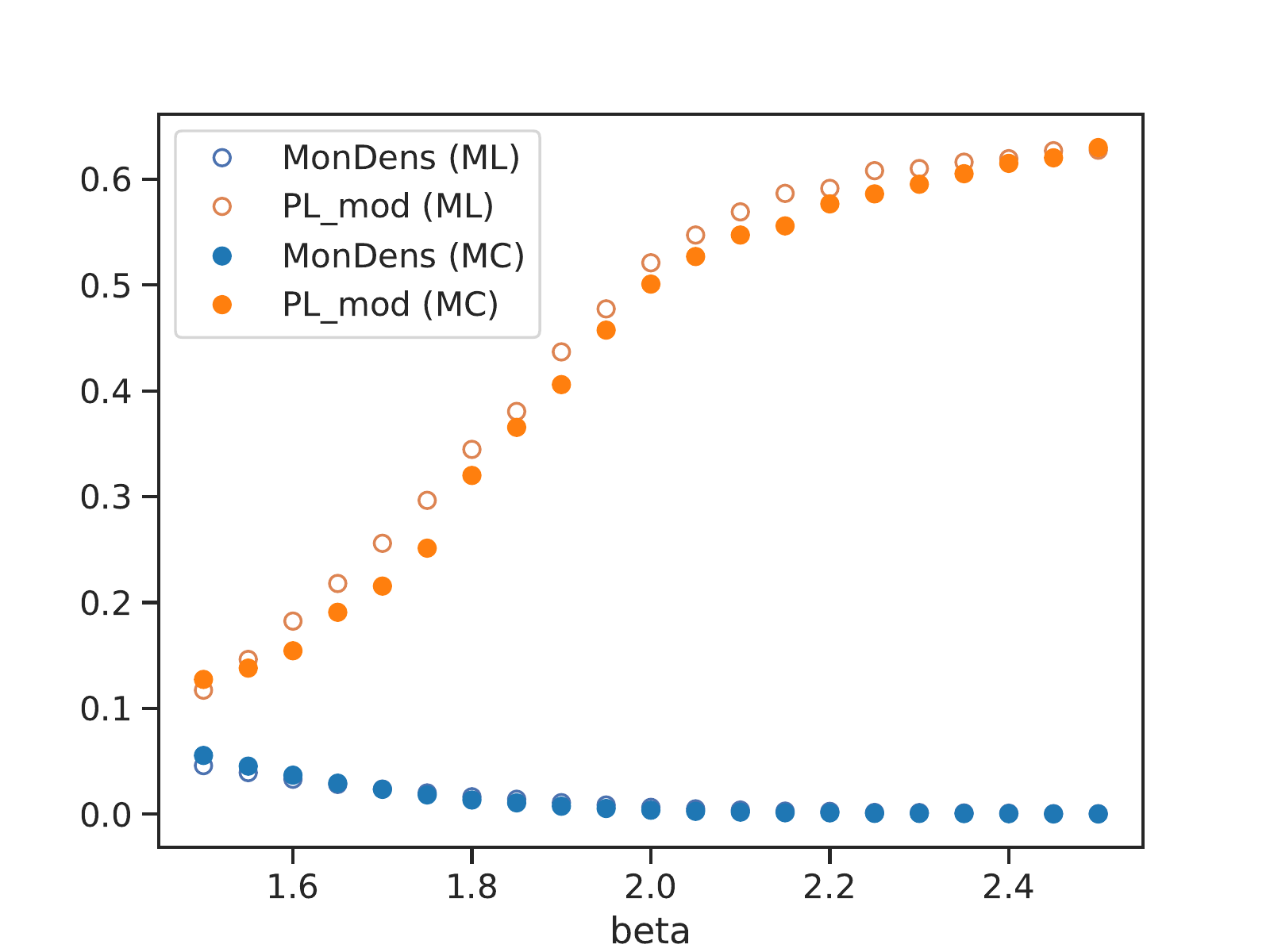}}
	\subcaptionbox{$4 \times 32^2$}{\includegraphics[scale=0.25]{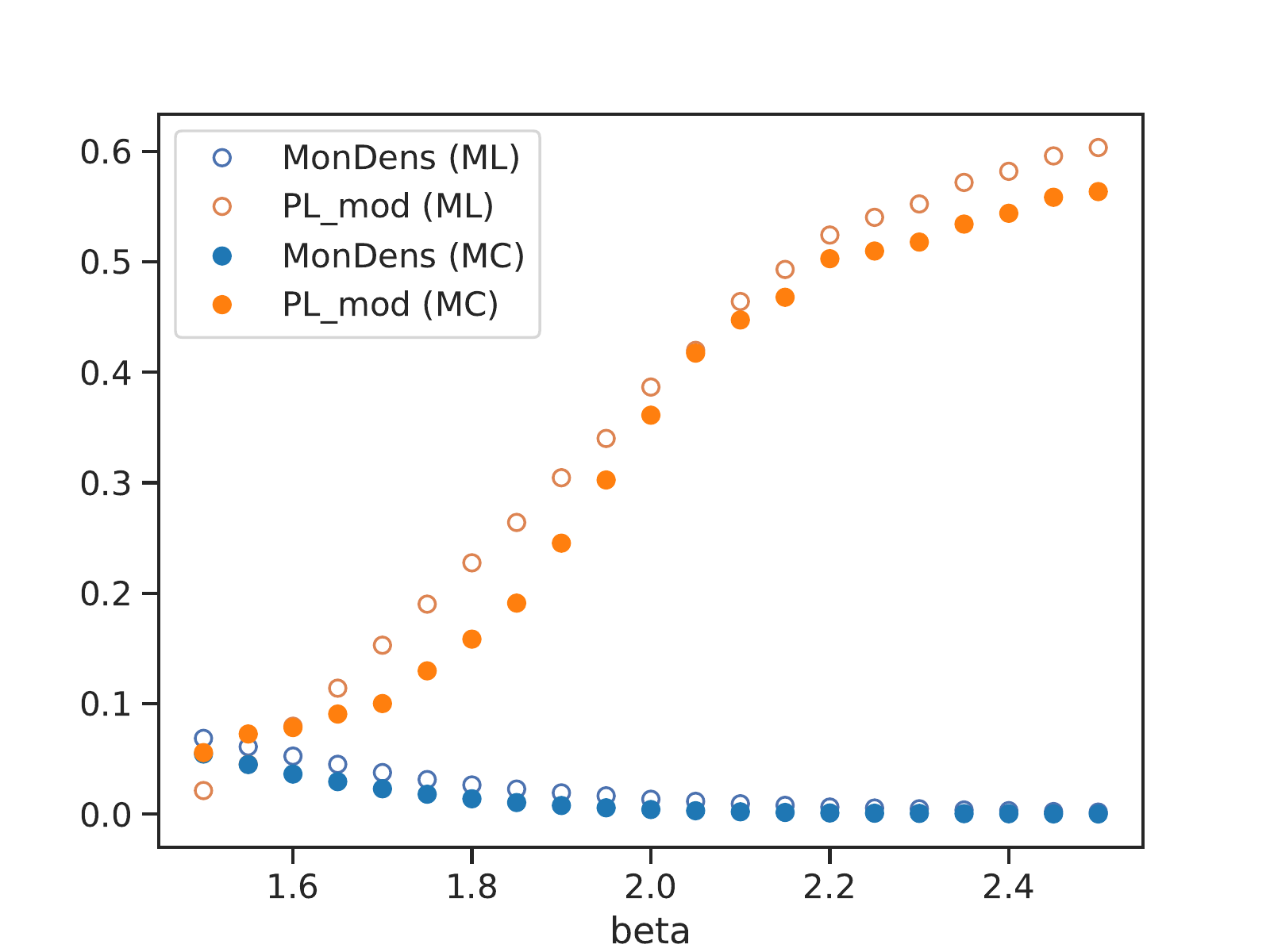}}
	\subcaptionbox{$6 \times 16^2$}{\includegraphics[scale=0.25]{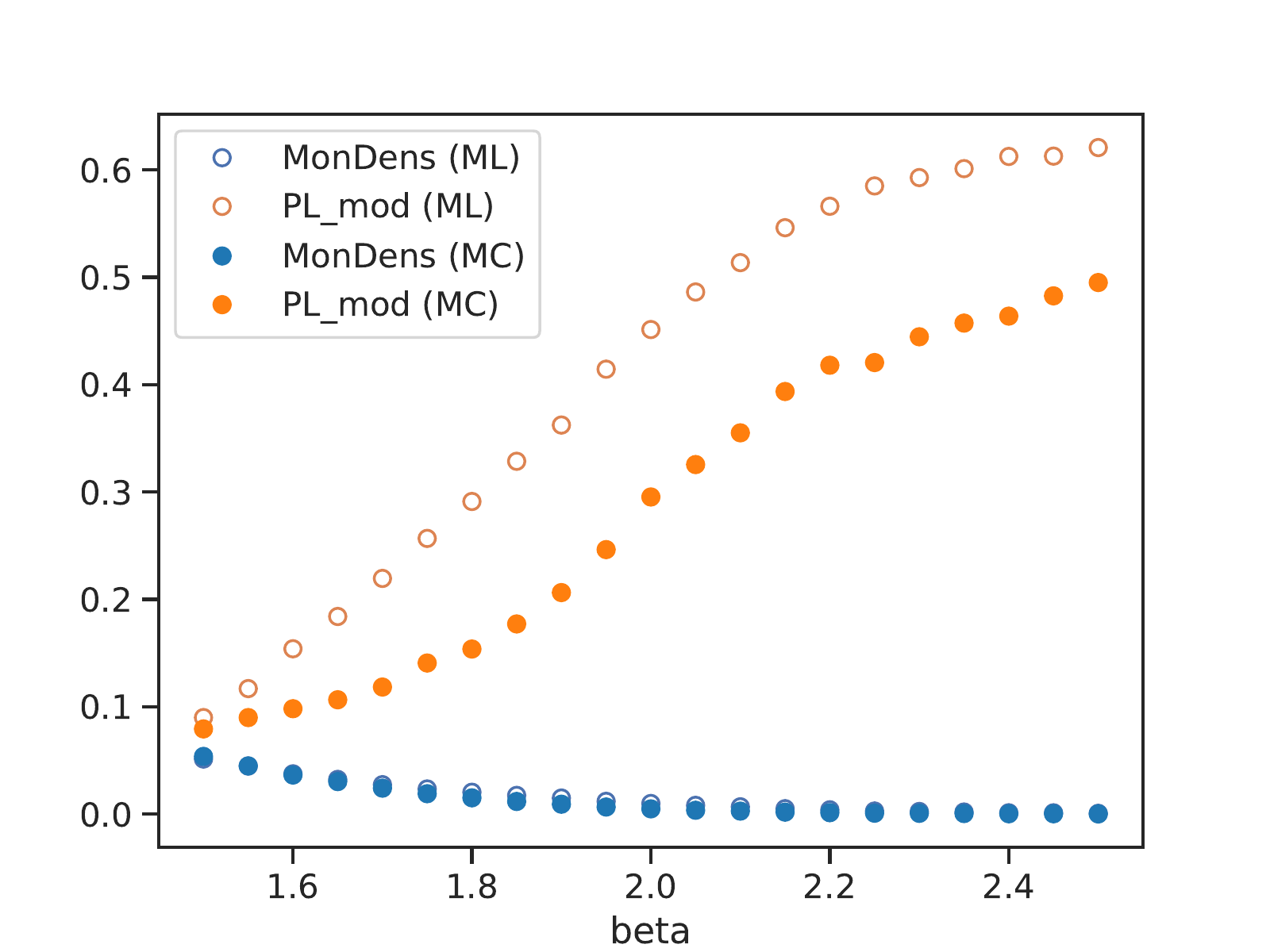}}
	\subcaptionbox{$6 \times 32^2$}{\includegraphics[scale=0.25]{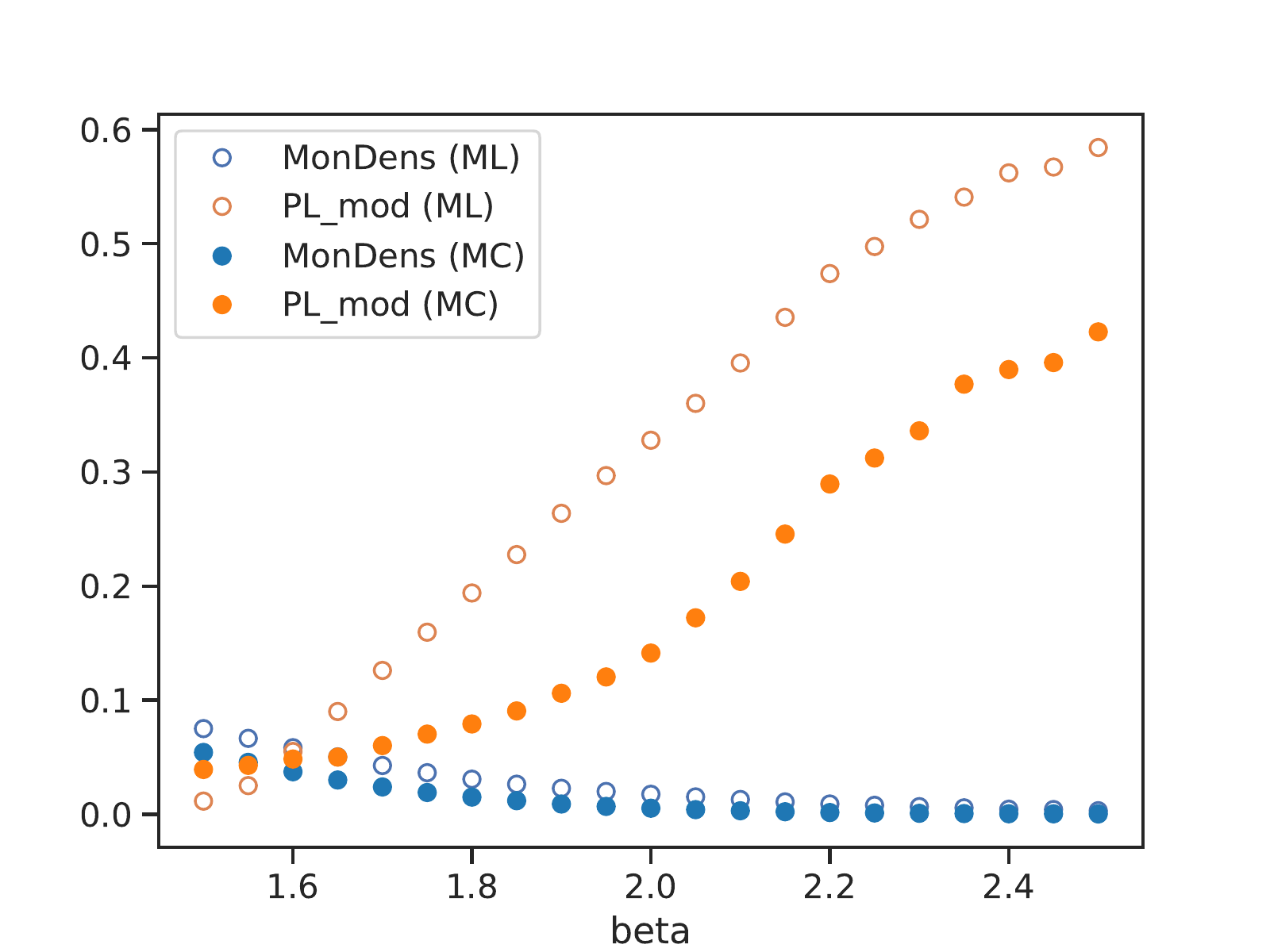}}
	\subcaptionbox{$8 \times 16^2$}{\includegraphics[scale=0.25]{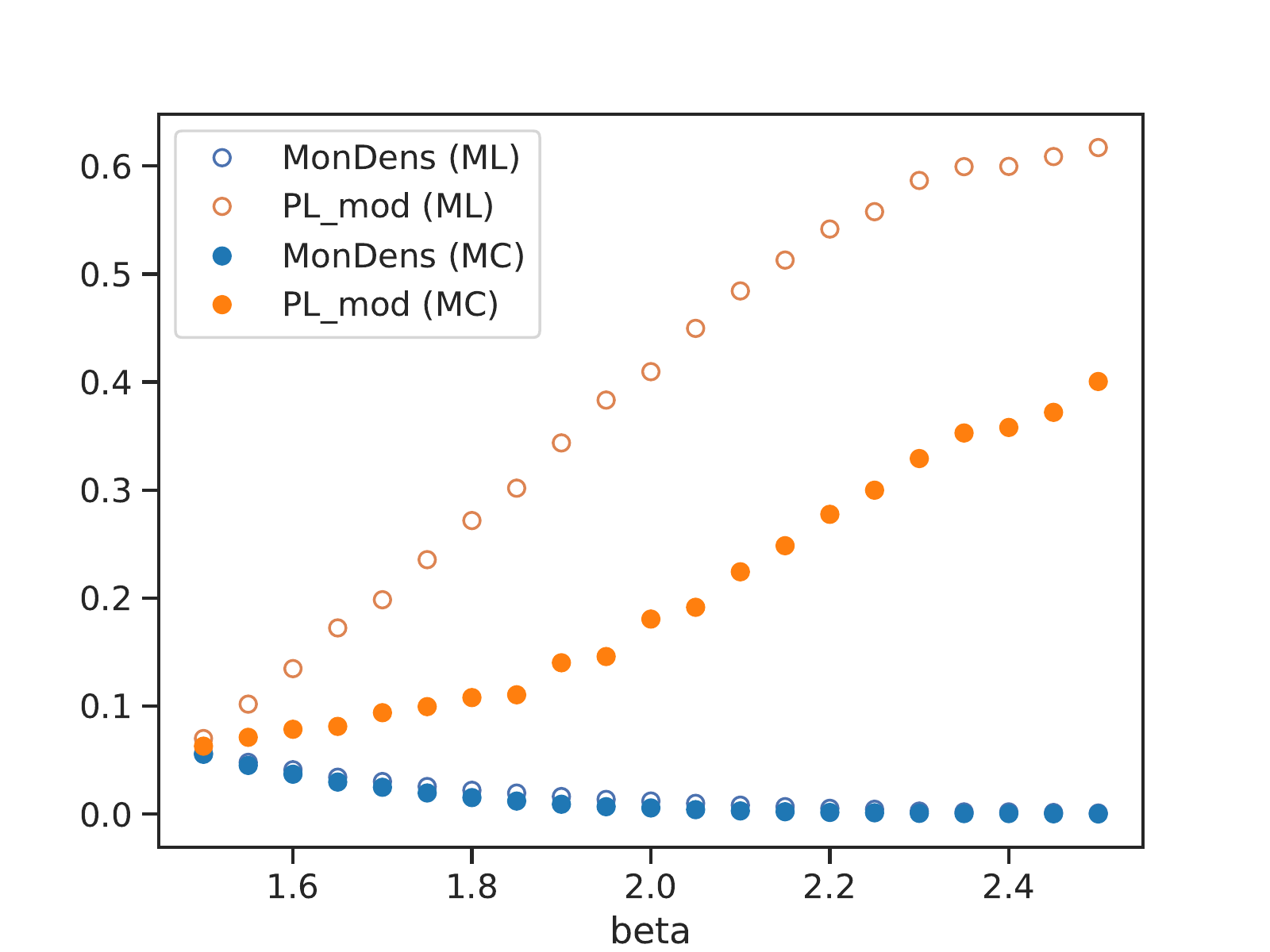}}
	\subcaptionbox{$8 \times 32^2$}{\includegraphics[scale=0.25]{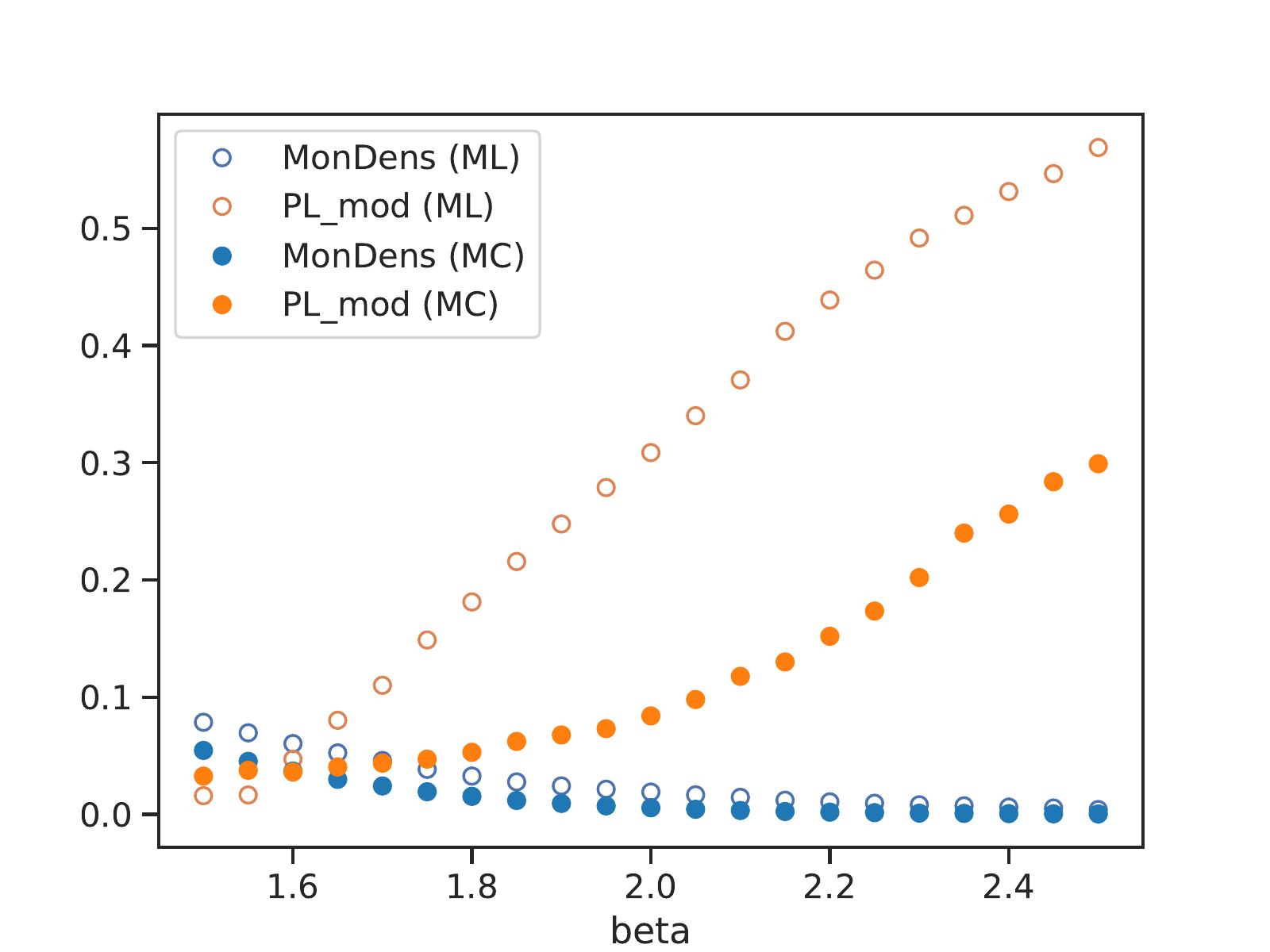}}
	\caption{\raggedright The monopole density $\mean{\rho}$ and the Polyakov loop $\mean{ L }$ in terms of $\beta$. The filled symbols show the actual quantities calculated with the help of Monte-Carlo (MC) technique and the open symbols are the predictions based on the machine learning (ML) of the monopole configurations.}
	\label{fig:plot-beta-L-rho}
\end{figure}

To better visualize the results, the joint predictions of $ L $ and $\phi$ in terms of the temperature are plotted in Figure~\ref{fig:temp-L-phi-4-16}.
Finally, the mean values $\mean{ L }_\beta$ and $\mean{\rho}_\beta$ in terms of the temperature are given in Figure~\ref{fig:plot-beta-L-rho}. Hereafter, we will use the notation ${\mathcal O}_\beta$ to stress the dependence of an operator ${\mathcal O}$ on the coupling constant~$\beta$. 

We observe from the different plots and from the performance measure that the different quantities are quite well learned by the neural network from the monopole configurations. The predictions for $p(\phi)$ and $ L $ turned out to be well correlated: there is some critical value of $ L $ for which the network predicts $\phi = 0$ for all configurations below, and and $\phi = 1$ above. Moreover, the network predictions have less variance than the real values.

\begin{table}[ht]
	\centering
%	\begin{tabular}{c|c}
%		& RMSE
%		\\
%		\hline
%		$ L $ & $\num{0.0893}$
%		\\
%		$\rho$ & $\num{0.00408}$
%		\\
%		$\beta$ & $\num{0.187}$
%		\\
%		$U$ & $\num{0.0166}$
%	\end{tabular}
	\begin{tabular}{c|c|c|c|c}
		     & $ L $       & $\rho$          &  $\beta$        &  $U$   \\
		\hline
		RMSE\ & $\ \num{0.0893}\ $  & $\ \num{0.00408}\ $  &  $\ \num{0.187}\ $  &  $\ \num{0.0166}\ $ 
	\end{tabular}
	\caption{\raggedright RMSE for the different continuous quantities.}
	\label{tab:rmse}
\end{table}

\begin{table}[ht]
	\centering
	\begin{tabular}{c|cc}
		$\phi$ & $p_c = 0.5$ & $p_c = 0.85$
		\\
		\hline
		accuracy & $\num{94.6}\%$ & $\num{92.6}\%$
		\\
		precision & $\num{95.1}\%$ & $\num{98.6}\%$
		\\
		recall & $\num{96.8}\%$ & $\num{90.1}\%$
%		\\
%		$F_1$ & $\num{0.960}$ & $\num{0.941}$
	\end{tabular}
	\caption{\raggedright %
		Performance of the phase classification quantified in terms of accuracy, precision and recall~\eq{eq:metrics}.
	}
	\label{tab:classif-metrics}
\end{table}

\begin{figure}[!htb]
	\centering
	\subcaptionbox{$6 \times 16^2$ -- real}{\includegraphics[scale=0.25]{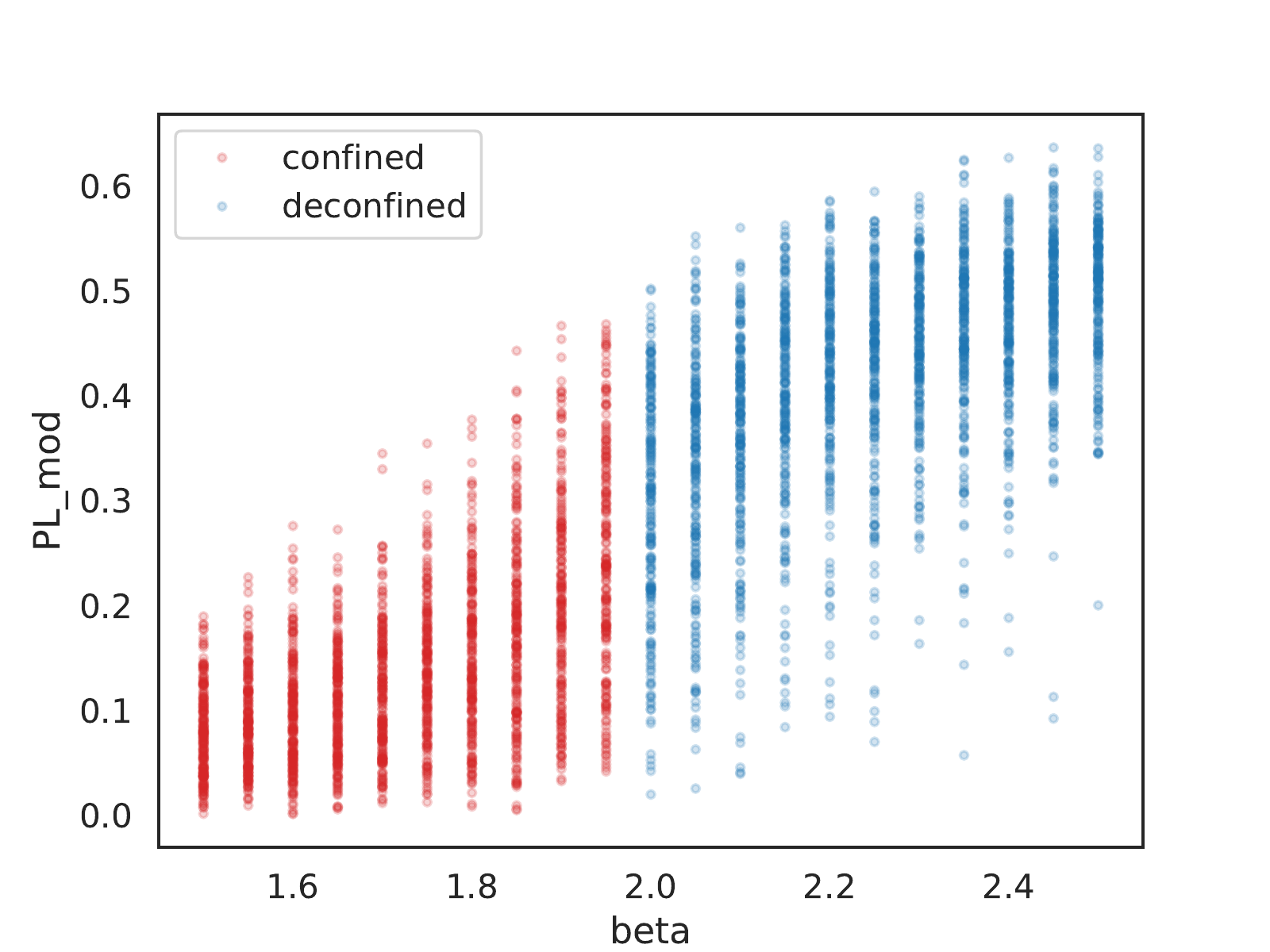}}
	\subcaptionbox{$6 \times 16^2$ -- predicted}{\includegraphics[scale=0.25]{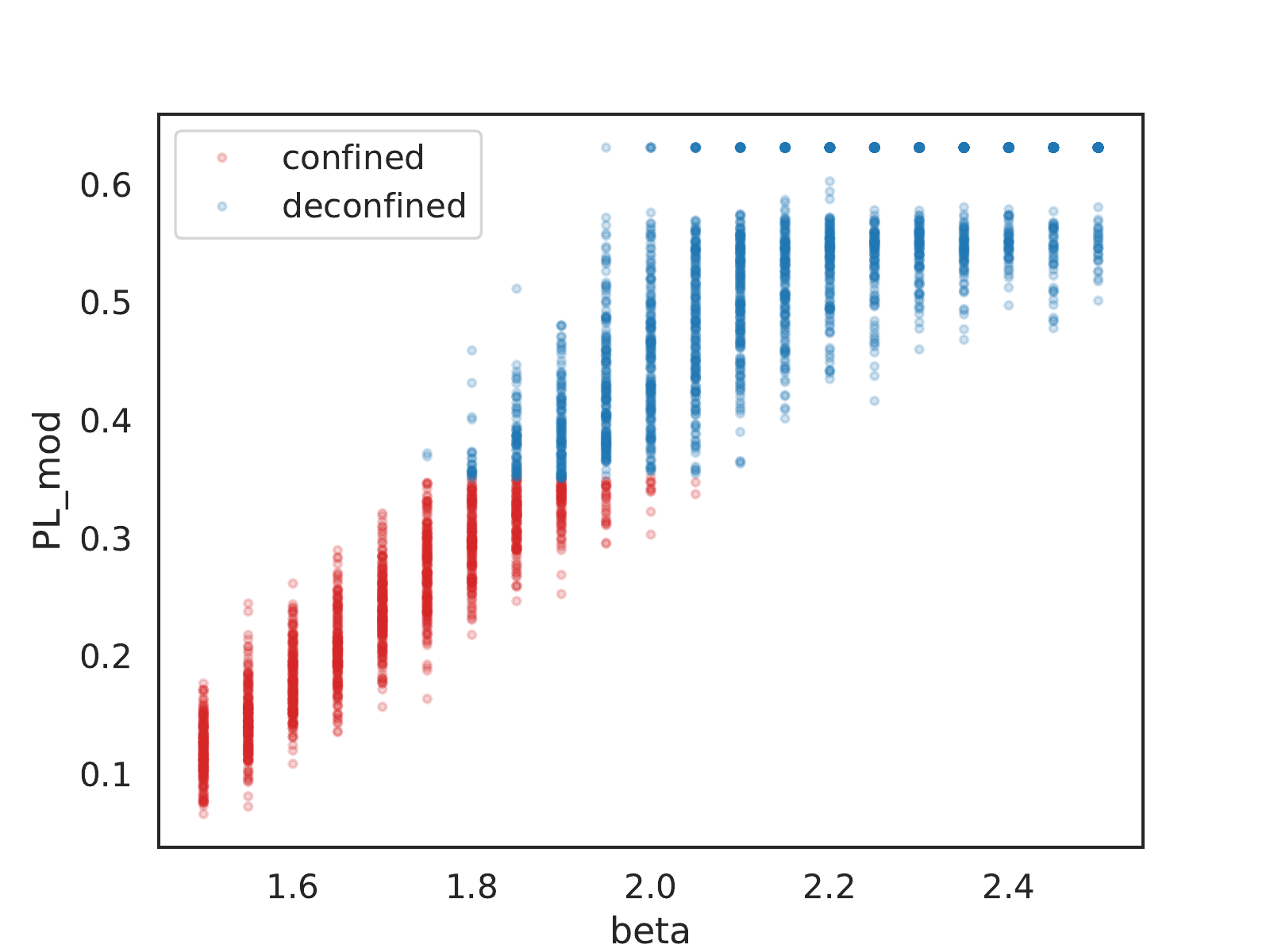}}
	\subcaptionbox{$8 \times 32^2$ -- real}{\includegraphics[scale=0.25]{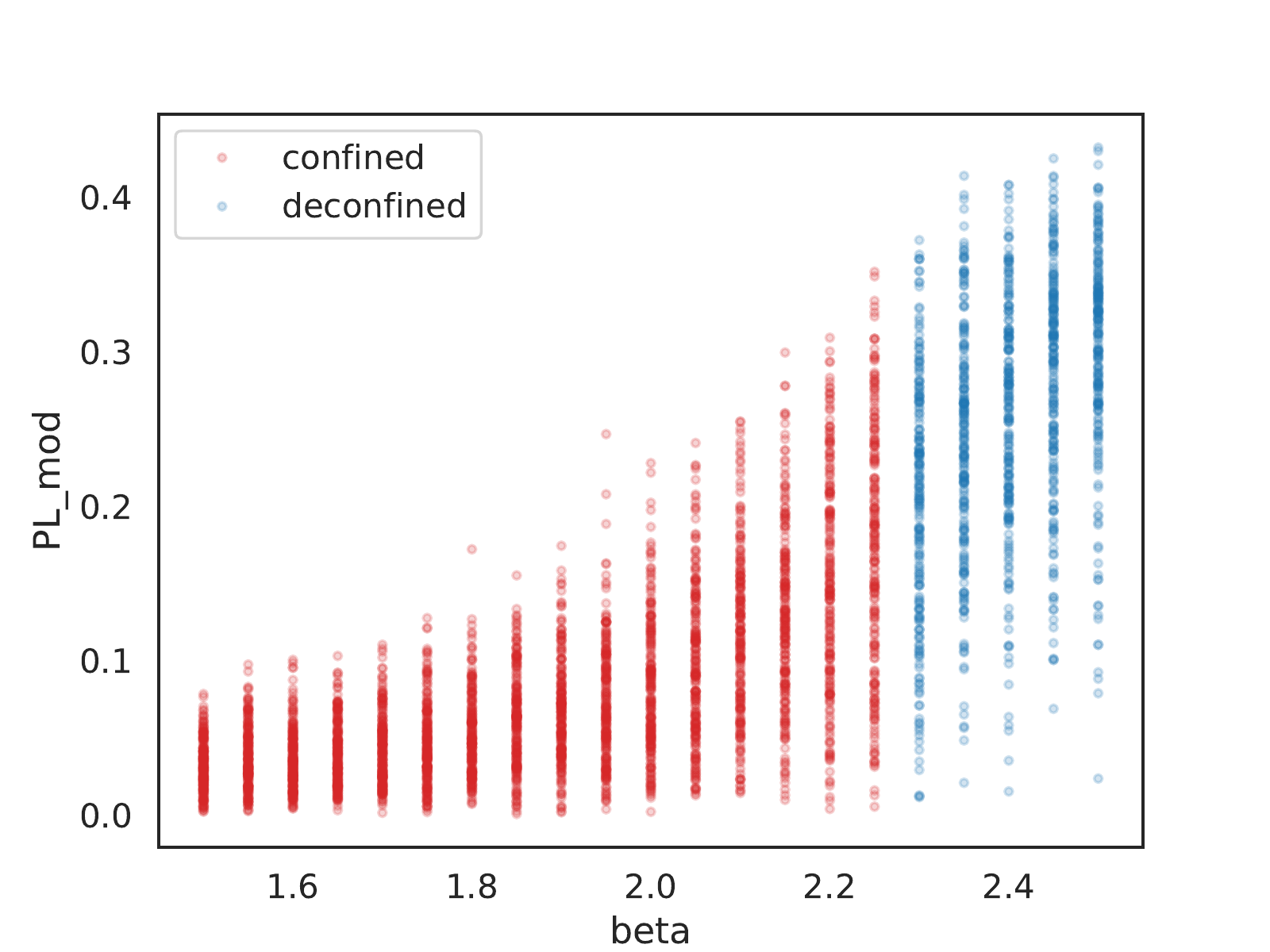}}
	\subcaptionbox{$8 \times 32^2$ -- predicted}{\includegraphics[scale=0.25]{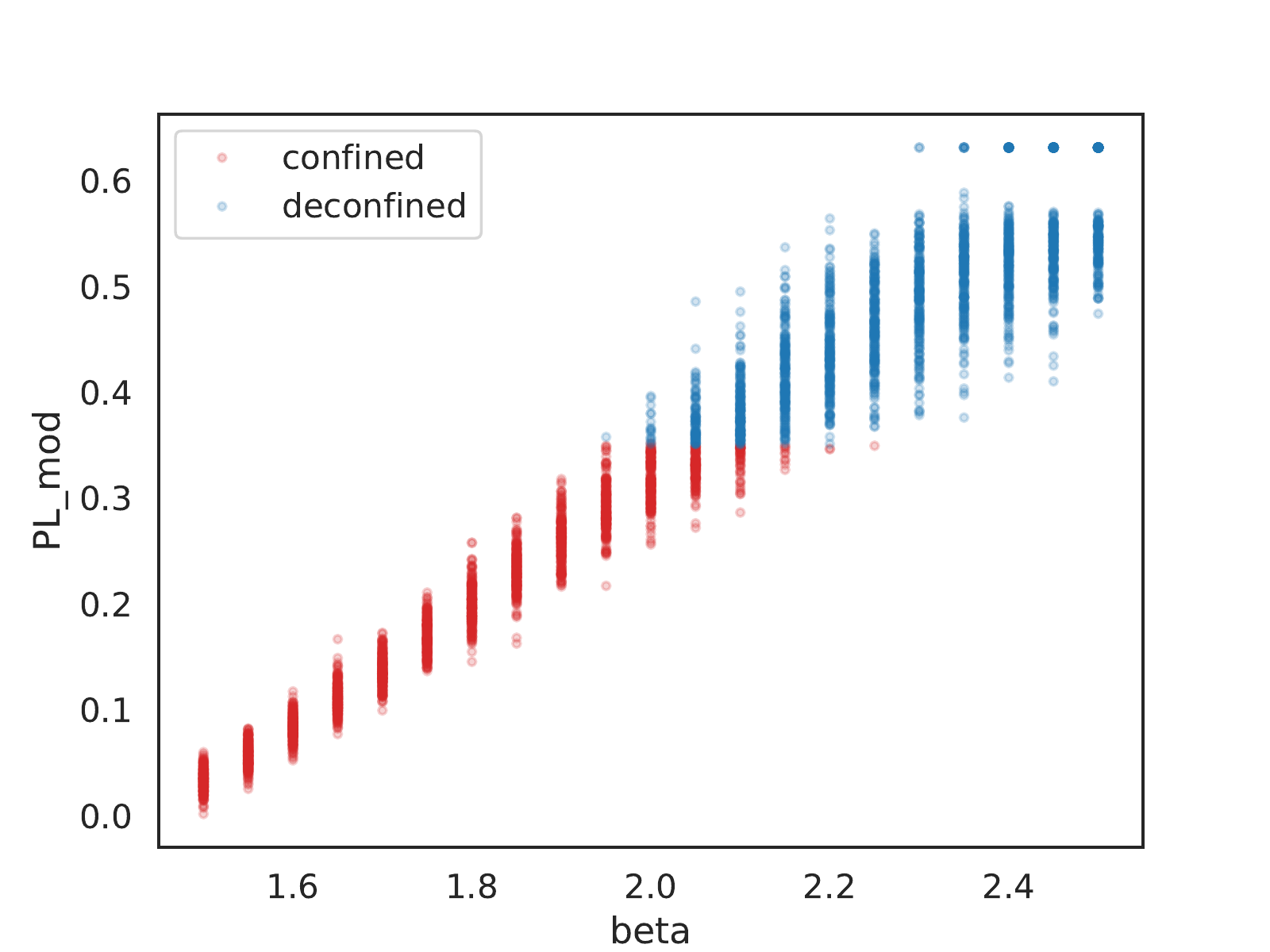}}
	\caption{\raggedright Examples of the distribution of the mean Polyakov loop $ L $ in terms of the coupling constant $\beta$ for each configuration, with the phase $\phi$ (red: $p = 0$, confined, blue: $p = 1$, deconfined), using $p_c = 0.5$.}
	\label{fig:plot-beta-L-phi-examples}
\end{figure}

We can then use the network to predict the different quantities at the other lattice sizes with $L_t = 4, 6, 8$ and $L_s = 16, 32$. The mean values $\mean{ L }_\beta$ and $\mean{\rho}_\beta$ in terms of $\beta$ are given in Figure~\ref{fig:plot-beta-L-rho}. Two examples of comparison of the values of $ L $ and $\phi$ in terms of the temperature are given in Figure~\ref{fig:plot-beta-L-phi-examples}. Notice that the actual value of the mean of the Polyakov loop is not important for the determination of the critical point: it is the maximum slope which is valid. Therefore the split of the original data (MC) and the predicted values (ML) does not play a crucial role.

\begin{figure}[!htb]
	\centering
	\subcaptionbox{$6 \times 16^2$}{\includegraphics[scale=0.25]{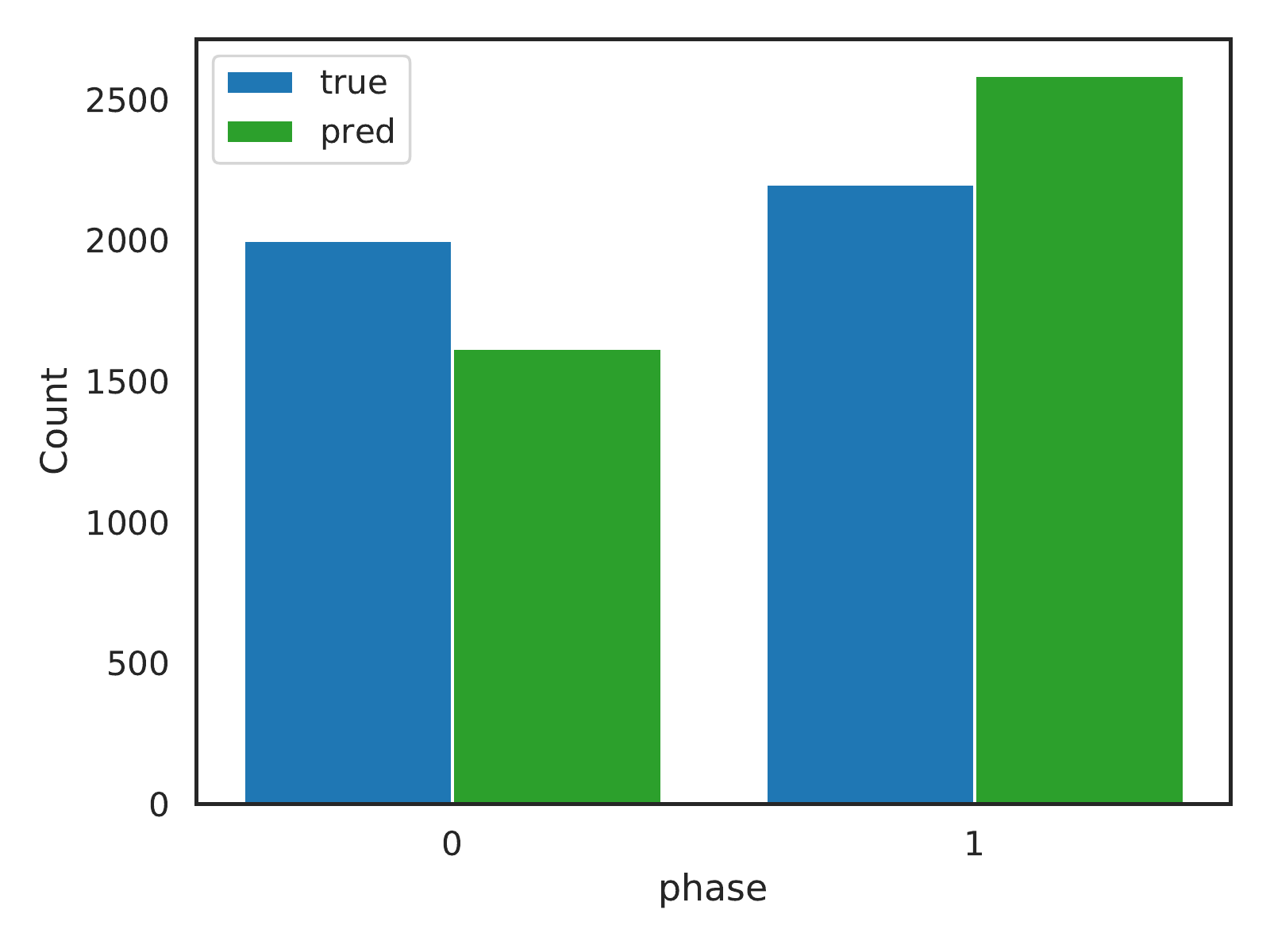}}
	\subcaptionbox{$8 \times 32^2$}{\includegraphics[scale=0.25]{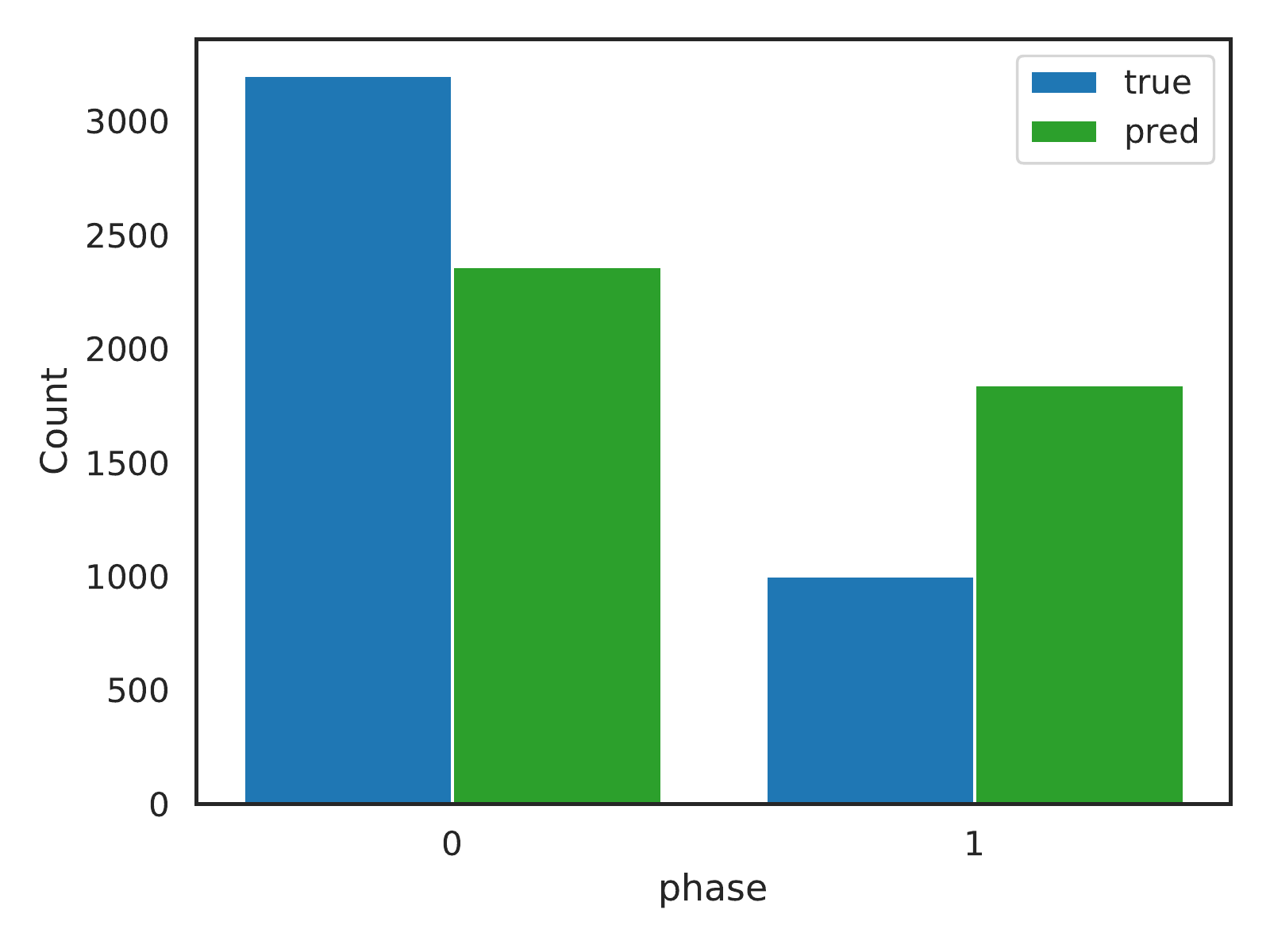}}
	\caption{\raggedright Examples of $\phi$ in terms of $\beta$ for each configuration, with the phase $\phi$ (red: $p = 0$, confined, blue: $p = 1$, deconfined), using $p_c = 0.5$.}
	\label{fig:plot-phase-distrib-examples}
\end{figure}

Finally, examples of the distribution of the phase prediction with $p_c = 0.5$ are given in Figure~\ref{fig:plot-phase-distrib-examples}.

While the monopole density prediction remains quite good, the errors in $ L $ and $\phi$ increase with the size of the lattice. The network makes predictions conservative towards the $4 \times 16^2$ lattice. The mean value $\mean{ L }_\beta$ is predicted to be linear on a larger and larger range of temperature, which prevents using the slope of the curve as a good indicator of the phase transition. For this reason, we will consider various ways to assess the temperature in the next section.

\subsection{Estimation of the critical temperature}

In this subsection, we evaluate the critical coupling $\beta_c$ and the associated critical temperature~\eq{eq:T:phys} from the quantities predicted by the neural networks. One may determine $\beta_c$ as a value of the coupling constant for which the slope of the Polyakov loop $\mean{ L }_\beta$ takes it maximum.
However, we pointed out in the previous section that the neural network does not see a sharp transition in terms of the Polyakov loop $ L $. On the other hand, the phase can provide a good estimation of the phase transition.

Indeed, we find that the neural networks have more and more difficulties for predicting the phase with certainty as one gets closer to the phase transition. To see this fact, we notice a substantial deviation of the values of $p(\phi)$ away from $0$ and $1$ as well as as a greater value of the variance. This observation in agreement with the ideas of Ref.~\cite{ref:phases:3} where it was found that the neural network should be more ``confused'' for predicting the phase close to the phase transition.

We consider five different methods to determine the critical coupling constant $\beta_c$:
\begin{enumerate}
	\item Maximum slope of the Polyakov loop $\mean{ L }_\beta$:
	\begin{equation}
		\beta_c
			= \argmax_{\beta} \pd_{\beta} \mean{ L }_\beta.
	\end{equation}

	\item Maximum uncertainty for the probability (the network predicts with equal chance the configuration to be in one phase or in the other):
	\begin{equation}
		\mean{p(\phi)}_\beta|_{\beta_c}
			= 0.5.
	\end{equation}

	\item Maximum variance of the probability:
	\begin{equation}
		\beta_c
			= \argmax_{\beta} \Var_{\beta}\big(p(\phi)\big).
	\end{equation}

	\item Maximum uncertainty for the phase:
	\begin{equation}
		\mean{\phi}_\beta{\biggl|}_{\beta = \beta_c}
			= 0.5.
	\end{equation}

	\item Maximum variance of the phase:
	\begin{equation}
		\beta_c
			= \argmax_{\beta} \Var_{\beta}(\phi).
	\end{equation}
\end{enumerate}
The difference between the methods 2) and 3) with 4) and 5) is that the former use the probability $p(\phi) \in [0, 1]$ (independent of $p_c$) while the latter use the phase label $\phi = 0, 1$ (which depends on $p_c$).
For each case, we also computed the temperature by first interpolating and then computing $\beta_c$, but this did not improve the results.

\onecolumngrid

\begin{table*}[!htp]
\centering
\begin{tabular}{c|cccccc}
& $4 \times 16^2$ & $4 \times 32^2$ & $6 \times 16^2$ & $6 \times 32^2$ & $8 \times 16^2$ & $8 \times 32^2$
\\
\hline
$ L $ slope & $1.85(1)$ & $2.02(5)$ & $1.90(1)$ & $2.12(4)$ & $1.96(7)$ & $2.06(12)$
\\
$\mean{p(\phi)}$ & $1.85(1)$ & $1.99(2)$ & $1.91(2)$ & $2.06(3)$ & $1.94(2)$ & $2.10(3)$
\\
$\Var p(\phi)$ & $1.83(3)$ & $1.96(3)$ & $1.88(3)$ & $2.04(2)$ & $1.91(3)$ & $2.07(2)$
% \\
% \hline
% mean (partial) & $1.84(1)$ & $1.99(3)$ & $1.90(2)$ & $2.07(3)$ & $1.94(3)$ & $2.08(4)$
\\
\hline
$\mean{\phi}$ ($p_c = 0.5$) & $1.84(2)$ & $1.98(2)$ & $1.90(2)$ & $2.05(3)$ & $1.94(2)$ & $2.08(3)$
\\
$\Var \phi$ ($p_c = 0.5$) & $1.81(2)$ & $1.95(2)$ & $1.89(2)$ & $2.02(2)$ & $1.90(2)$ & $2.06(3)$
\\
\hline
$\mean{\phi}$ ($p_c = 0.85$) & $1.91(2)$ & $2.08(3)$ & $2.00(2)$ & $2.17(3)$ & $2.05(2)$ & $2.21(5)$
\\
$\Var \phi$ ($p_c = 0.85$) & $1.90(2)$ & $2.06(3)$ & $1.98(3)$ & $2.14(3)$ & $2.02(3)$ & $2.18(5)$
% \\
% \hline
% mean & $1.85(1)$ & $1.98(2)$ & $1.90(1)$ & $2.04(2)$ & $1.94(2)$ & $2.06(3)$
\\
\hline
MC & 1.81 & 1.93 & 1.98 & 2.14 & 2.10 & 2.29
\end{tabular}
\caption{\raggedright %
	Predictions of the critical coupling $\beta_c$ given by the neural network using different methods and quantities (mean values and standard deviation by averaging over $10$ neural networks). The first three rows do not depend on the choice of the probability threshold $p_c$ of the decision function~\eq{eq:dec-fn}.}
\label{tab:pred-betac}
\end{table*}

\begin{table*}[!htp]
\centering
\begin{tabular}{c|cccccc}
& $4 \times 16^2$ & $4 \times 32^2$ & $6 \times 16^2$ & $6 \times 32^2$ & $8 \times 16^2$ & $8 \times 32^2$
\\
\hline
$ L $ slope & $2.21(1)\%$ & $4.67(237)\%$ & $4.04(1)\%$ & $1.64(176)\%$ & $6.66(350)\%$ & $10.05(524)\%$
\\
$\mean{p(\phi)}$ & $2.49(83)\%$ & $3.11(104)\%$ & $3.28(115)\%$ & $3.74(140)\%$ & $7.62(95)\%$ & $8.52(118)\%$
\\
$\Var p(\phi)$ & $1.38(83)\%$ & $1.82(166)\%$ & $5.05(168)\%$ & $4.91(107)\%$ & $8.81(152)\%$ & $9.61(107)\%$
% \\
% \hline
% mean (partial) & $1.84(61)\%$ & $3.20(135)\%$ & $4.12(88)\%$ & $3.27(138)\%$ & $7.70(151)\%$ & $9.39(167)\%$
\\
\hline
$\mean{\phi}$ ($p_c = 0.5$) & $1.88(66)\%$ & $2.59(127)\%$ & $3.79(76)\%$ & $4.21(148)\%$ & $7.62(95)\%$ & $9.39(109)\%$
\\
$\Var \phi$ ($p_c = 0.5$) & $0.72(50)\%$ & $1.35(78)\%$ & $4.55(101)\%$ & $5.61(114)\%$ & $9.28(71)\%$ & $10.26(117)\%$
\\
\hline
$\mean{\phi}$ ($p_c = 0.85$) & $5.52(110)\%$ & $7.77(172)\%$ & $1.31(76)\%$ & $1.54(110)\%$ & $2.38(106)\%$ & $3.58(199)\%$
\\
$\Var \phi$ ($p_c = 0.85$) & $4.97(123)\%$ & $6.48(139)\%$ & $1.52(72)\%$ & $1.12(84)\%$ & $3.57(160)\%$ & $4.58(196)\%$
% \\
% \hline
% mean & $2.27(47)\%$ & $2.40(116)\%$ & $3.91(71)\%$ & $4.46(93)\%$ & $7.65(104)\%$ & $10.26(117)\%$
\end{tabular}
\caption{\raggedright %
	Relative errors of the ML prediction, $\abs{\beta_c^{\text{ML}} - \beta_c^{\text{MC}}} / \beta_c^{\text{MC}}$, using different methods (mean values and standard deviation by averaging over $10$ neural networks). The first three rows do not depend on $p_c$.
}
\label{tab:errors-betac}
\end{table*}

\twocolumngrid

\begin{figure}[htp]
	\centering
	\subcaptionbox{\label{fig:mean-phase-prob}}{\includegraphics[scale=0.5]{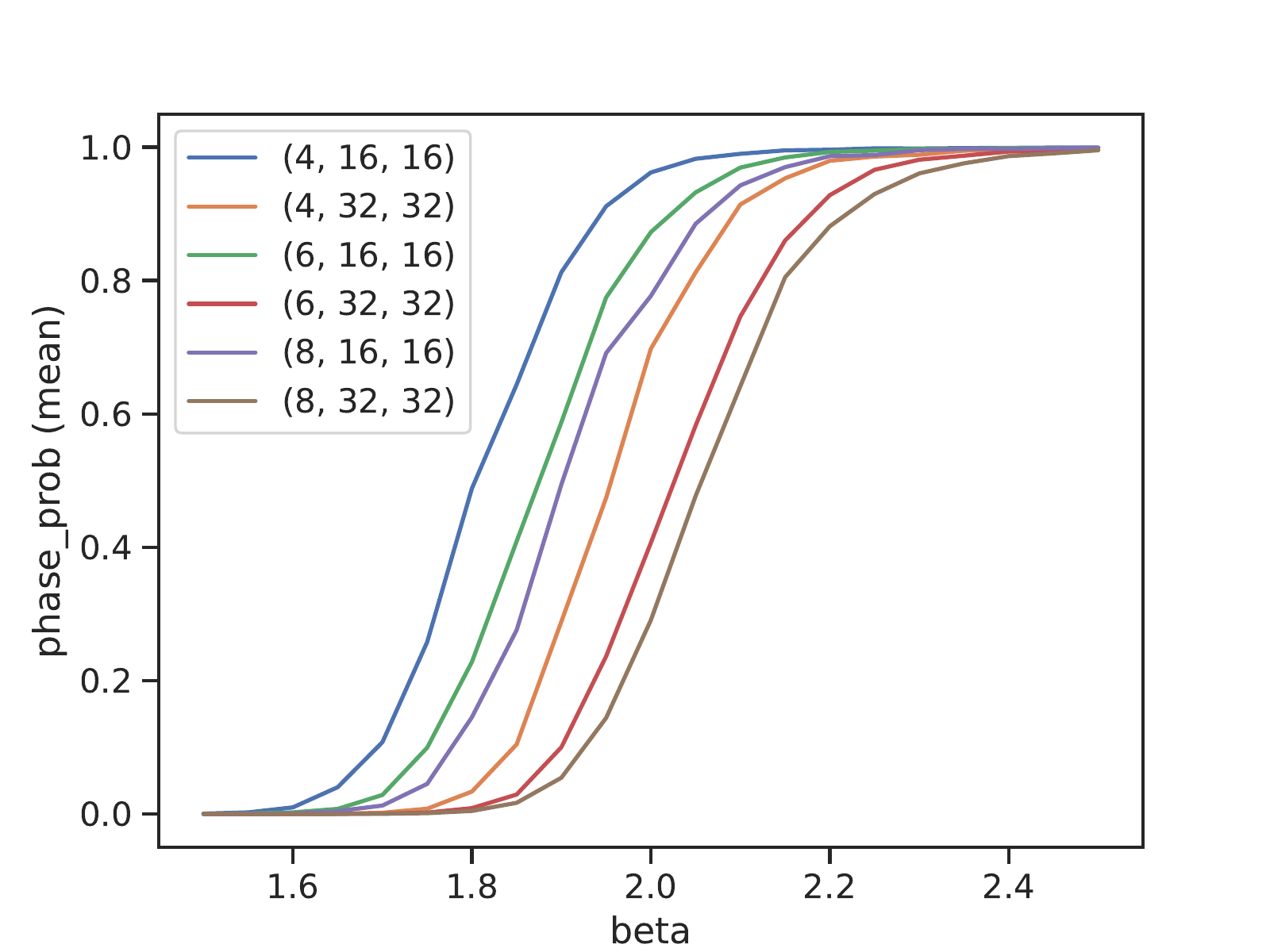}}
	\subcaptionbox{\label{fig:var-phase-prob}}{\includegraphics[scale=0.5]{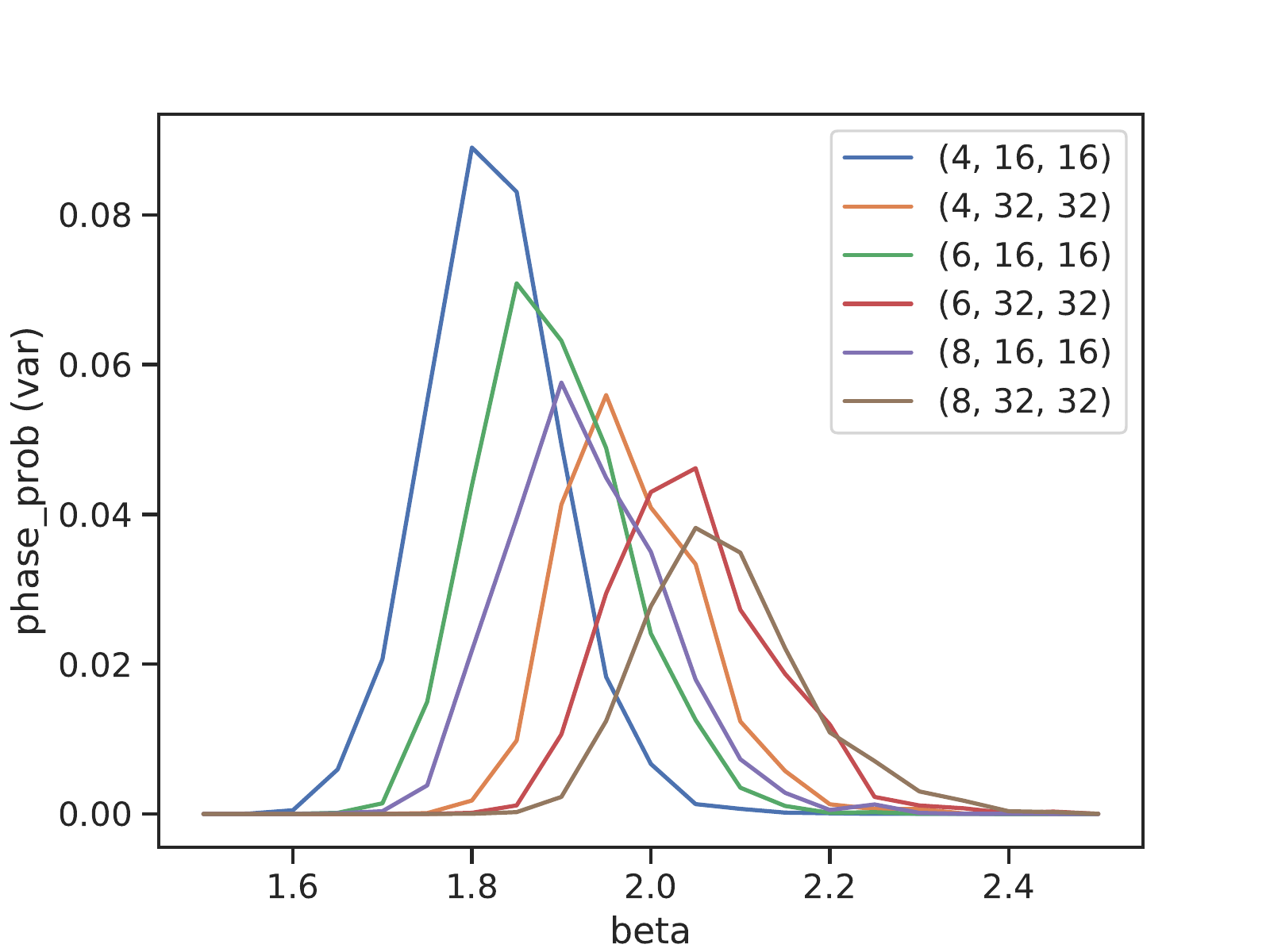}}
	\caption{\raggedright (a) Mean $\mean{p(\phi)}$ and (b) variance $\Var p(\phi)$ of the distribution $p(\phi)$ for different lattice geometries.}
	\label{fig:stat-phase-prob}
\end{figure}

The results for the critical coupling $\beta_c$ and the relative errors in its determination $\abs{\beta_c^{\text{ML}} - \beta_c^{\text{MC}}} / \beta_c^{\text{MC}}$ are given in Tables~\ref{tab:pred-betac} and~\ref{tab:errors-betac}, respectively. The tables indicate the mean values and standard deviations obtained by training $n = 10$ different models.
The mean averages and the variances of the distribution $p(\phi)$ are given in Figures~\ref{fig:mean-phase-prob} and~\ref{fig:var-phase-prob}.

Various methods provide slightly different predictions for the critical coupling constant~$\beta_c$ which generally lie within a few percent (10\% in the worst case) from the actual position of the transition.

\begin{figure}[!htb]
	\centering
	\includegraphics[scale=0.4]{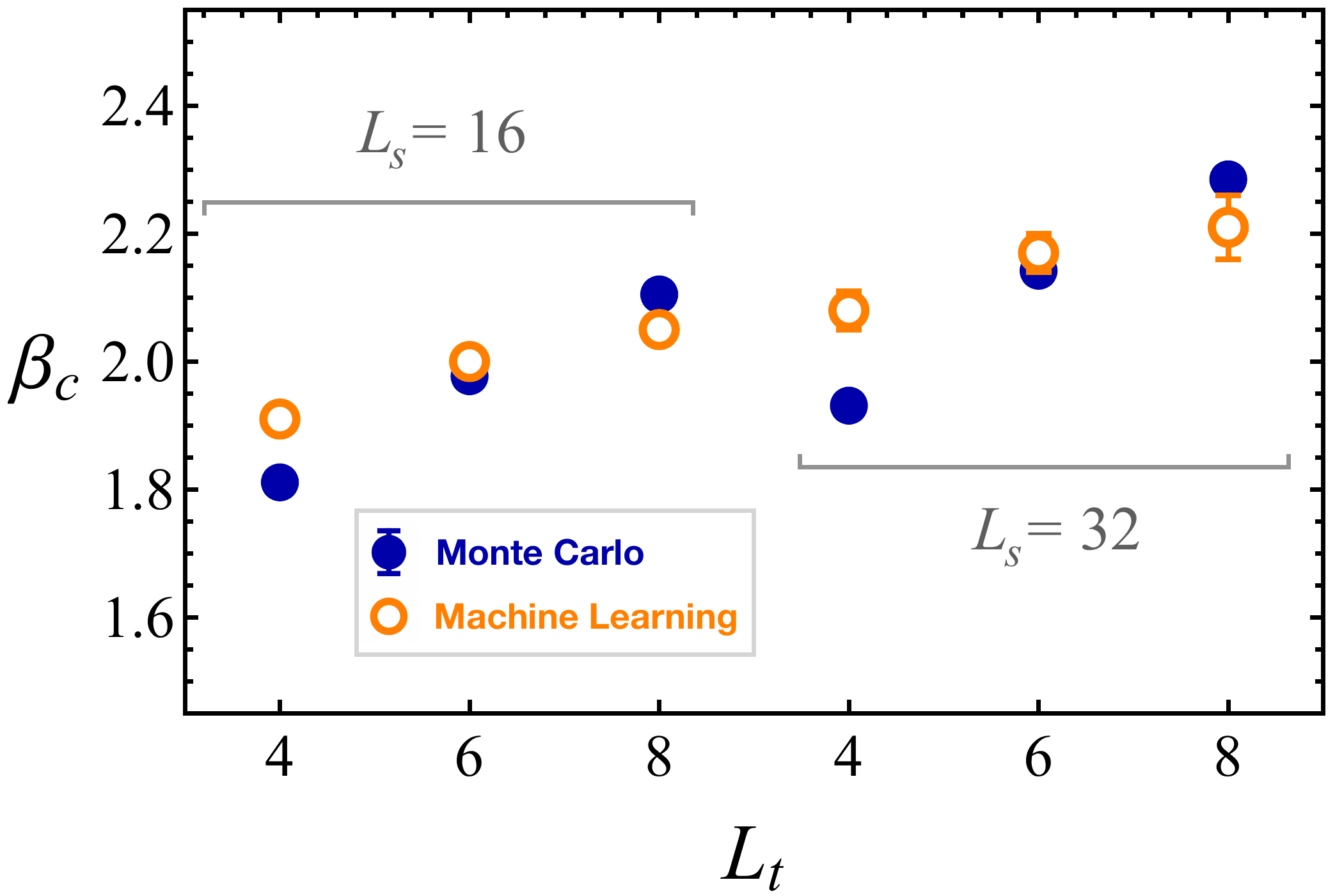}
	\caption{\raggedright The critical coupling constant $\beta_c$ of compact electrodynamics at various lattices $L_s^2 \times L_t$. The full symbols: the Monte-Carlo results obtained from the original gauge-field configurations. The open symbols: the prediction of the neural network based on monopole configurations only.}
	\label{fig:betas}
\end{figure}

In Figure~\ref{fig:betas} we show the critical coupling $\beta_c$ of the deconfinement phase transition obtained with the help of the Monte-Carlo estimation at the original gauge-field configurations. We compare these numbers with the prediction of the neural network (ML) using the monopole configurations only. For illustration, we use the estimation based on the maximal uncertainty (``the network's confusion'') of the phase label $\phi$ with the threshold $\phi_c = 0.85$. The other criteria listed in Table~\ref{tab:pred-betac} give similar predictions but with globally lower accuracy according to Table~\ref{tab:errors-betac}.

While we notice that the results are quite close to each other, the biggest mismatch comes from the lattices with the smallest temporal extension $L_t = 4$. At larger sizes $L_t = 6$ and $L_t = 8$, which are closer to thermodynamic limit, the agreement between the real (MC) and the predicted (ML) values is much closer. We suggest that this mismatch appears because the network cannot distinguish between isolated monopoles and antimonopoles at one side and monopole-antimonopole pairs bound via the periodic boundary at the other side. This effect naturally overestimates the density of the unbound monopoles and gives an overestimated prediction of the deconfining coupling (temperature) $\beta$, as it is seen in Figure~\ref{fig:betas}. This unwanted effect disappears closer to the thermodynamic limit at larger temporal extensions $L_t$. 

Finishing this section, we would like to comment on the errors of our approach. We see that the $p_c$-independent and $p_c = 0.5$ methods give results with similar errors. The latter grow with the size of the lattice.

There exist different possibilities for mitigating the errors due to the extrapolation at higher lattice sizes.
A first possibility is to change the probability threshold $p_c$ in the decision function \eqref{eq:dec-fn}. We found that a probability threshold is $p_c = 0.85$ give much better results. The reason for this improvement is that the neural network is conservative towards the results for $L_t = 4$ and $L_s = 16$ where $\beta_c$ is lower. Increasing $p_c$ pushes the transition further as it moves more configurations in the confined phase. However, it could be necessary to increase further the values when extending at even larger lattices. Another possibility would be to find a function $p_c = p_c(L_t, L_s)$ by considering the results on few lattices (this method is interesting when one wants to make predictions for a number of lattices much higher than the one used for training: in that case, it is fine to ``lose'' some lattices for training without reducing much the predictive power).

\begin{figure}[htp]
	\centering
	\includegraphics[scale=0.5]{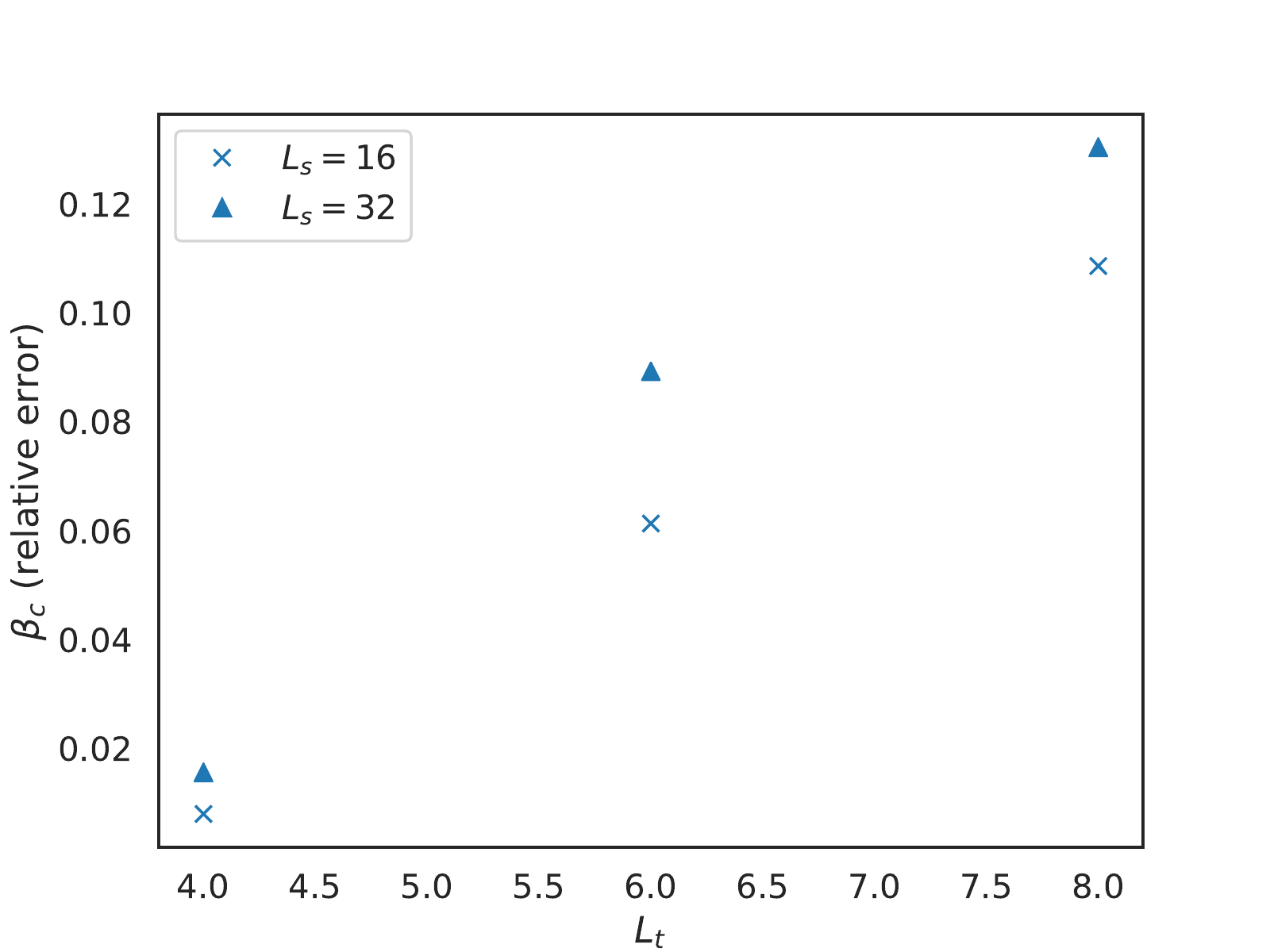}
	\caption{\raggedright Relative errors in terms of $L_t$ and $L_s$ for $\beta_c$ computed from $\mean{p(\phi)}$.}
	\label{fig:rel-error-Lt-Ls}
\end{figure}

Another possibility to reduce the errors is to find a pattern in the errors made in the predictions of $\beta_c$. In fact, one finds that the relative error grows linearly with $L_t$ (Figure~\ref{fig:rel-error-Lt-Ls}). This observation could in principle be used to correct the predictions if one knows the correct result for few lattices.

In both cases, a proper analysis would require to extend the learning process to different lattice sizes that can be understood as a form of boosting (a ML technique to correct iteratively a result). This analysis goes beyond the scope of the present paper which focuses on what can be learned by training a neural network on a single lattice size.

\section{Conclusions}

We applied the machine learning techniques to investigate the phase transition produced by the dynamics of topological defects. We used the compact U(1) gauge theory in three spacetime dimensions, which exhibits the deconfining phase transition associated with the binding of the Abelian monopoles at a critical temperature. The system goes from the monopole gas at low temperature to a gas of monopole-antimonopole pairs at high temperature through an infinite-order phase transition of the Berezinskii--Kosterlitz--Thouless type.

The neural network uses the supervised learning technique to acquire knowledge about monopole configurations generated by the standard Monte-Carlo technique. The network processes the monopole configurations as holograms (three-dimensional images) and studies how to associate these monopole holograms with the vacuum expectation value of the Polyakov loop (the order parameter of the transition) at relatively small lattices.

After completion of the training stage, the neural network uses the monopole configurations at larger-volume lattices to distinguish confinement and deconfinement phases, determine the deconfinement transition point, and predict monopole densities as well as the expectation values of the Polyakov loop. 

We show that the model can determine the transition temperature with reasonably good accuracy, which depends on the criteria implemented in the algorithm. In agreement with Ref.~\cite{ref:phases:3}, we found that the best criterion for locating the phase transition corresponds to the degree of the confusion experienced by the neural network engaged with the task to determine the transition point. The maximum confusion appears in the close vicinity of the transition, seen via the enhanced variance of the probability of finding a definite phase.

Expectedly, the neural network is successful in the prediction of the mean monopole density. While the predicted Polyakov loop differs from the behavior of the original order parameter, the critical inflection points of both quantities are close to each other.

We conclude that the neural network can see the position of the deconfining phase transition -- using the maximum confusion as reliable criterion --  sensing the transition via the holograms of magnetic monopoles. The neural network correctly addresses the thermodynamic bulk properties being able to extrapolate its predictions to lattices with different volumes.

\acknowledgments

We would like to thank Mohamed El Amine Seddik for useful discussions on neural networks. The research was partially supported by a grant of the Russian Foundation for Basic Research No. 18-02-40121 mega and by Grant No. 0657-2020-0015 of the Ministry of Science and Higher Education of Russia. The numerical simulations were performed at the computing cluster Vostok-1 of Far Eastern Federal University. H.E. was supported by a Carl Friedrich von Siemens Research Fellowship of the Alexander von Humboldt Foundation during part of this project.
The work of H.E. is partially supported by the MIUR PRIN Contract 2015MP2CX4 “Non-perturbative Aspects Of Gauge Theories And Strings”.


\begin{thebibliography}{99}

\bibitem{Polyakov:1976fu}
  A.~M.~Polyakov,
  ``Quark Confinement and Topology of Gauge Groups,''
  Nucl.\ Phys.\ B {\bf 120}, 429 (1977).
%  doi:10.1016/0550-3213(77)90086-4
  %%CITATION = doi:10.1016/0550-3213(77)90086-4;%%

\bibitem{Gross:1980br}
D.~J.~Gross, R.~D.~Pisarski and L.~G.~Yaffe,
``QCD and Instantons at Finite Temperature,''
Rev. Mod. Phys. \textbf{53}, 43 (1981).
%doi:10.1103/RevModPhys.53.43

\bibitem{Fiebig:1990uh}
H.~Fiebig and R.~Woloshyn,
``Monopoles and chiral symmetry breaking in lattice QED in three-dimensions,''
Phys. Rev. D \textbf{42}, 3520-3523 (1990).
%doi:10.1103/PhysRevD.42.3520


\bibitem{ref:book:Herbut}
I. Herbut,
``A Modern Approach to Critical Phenomena'' (Cambridge University Press, 2007).

\bibitem{ref:book:Kleinert}
Hagen Kleinert,
Multivalued Fields: In Condensed Matter, Electromagnetism, and Gravitation (World Scientific, 2008, Singapore).


\bibitem{Chernodub:2017mhi}
M.~Chernodub, V.~Goy and A.~Molochkov,
``Nonperturbative Casimir effect and monopoles: compact Abelian gauge theory in two spatial dimensions,''
Phys. Rev. D \textbf{95}, no.7, 074511 (2017).
%doi:10.1103/PhysRevD.95.074511
%[arXiv:1703.03439 [hep-lat]].
%10 citations counted in INSPIRE as of 09 Jun 2020

\bibitem{Chernodub:2018pmt}
M.~Chernodub, V.~Goy, A.~Molochkov and H.~H.~Nguyen,
``Casimir Effect in Yang-Mills Theory in D=2+1,''
Phys. Rev. Lett. \textbf{121}, no.19, 191601 (2018).
%doi:10.1103/PhysRevLett.121.191601
%[arXiv:1805.11887 [hep-lat]].

\bibitem{Chernodub:2019kon}
M.~Chernodub, H.~Erbin, I.~Grishmanovskii, V.~Goy and A.~Molochkov,
``Casimir effect with machine learning,''
[arXiv:1911.07571 [hep-lat]].

\bibitem{Parga:1981tm}
N.~Parga,
``Finite Temperature Behavior of Topological Excitations in Lattice Compact QED,''
Phys. Lett. B \textbf{107}, 442 (1981)
%doi:10.1016/0370-2693(81)91225-9

\bibitem{Coddington:1986jk}
P.~D.~Coddington, A.~J.~Hey, A.~Middleton and J.~S.~Townsend,
``The Deconfining Transition for Finite Temperature U(1) Lattice Gauge Theory in (2+1)-dimensions,''
Phys. Lett. B \textbf{175}, 64-68 (1986)
%doi:10.1016/0370-2693(86)90332-1ISTEXISTEX

\bibitem{Chernodub:2001ws}
M.~Chernodub, E.~M.~Ilgenfritz and A.~Schiller,
``A Lattice study of 3-D compact QED at finite temperature,''
Phys. Rev. D \textbf{64}, 054507 (2001)
%doi:10.1103/PhysRevD.64.054507
%[arXiv:hep-lat/0105021 [hep-lat]].
%33 citations counted in INSPIRE as of 09 Jun 2020

\bibitem{Borisenko:2008sc}
O.~Borisenko, M.~Gravina and A.~Papa,
``Critical behavior of the compact 3d U(1) theory in the limit of zero spatial coupling,''
J. Stat. Mech. \textbf{0808}, P08009 (2008).
%doi:10.1088/1742-5468/2008/08/P08009ISTEXISTEX
%[arXiv:0806.2081 [hep-lat]].

\bibitem{Borisenko:2010qe}
O.~Borisenko, R.~Fiore, M.~Gravina and A.~Papa,
``Critical behavior of the compact 3d U(1) gauge theory on isotropic lattices,''
J. Stat. Mech. \textbf{1004}, P04015 (2010).
%doi:10.1088/1742-5468/2010/04/P04015ISTEXISTEX
%[arXiv:1001.4979 [hep-lat]].
%13 citations counted in INSPIRE as of 09 Jun 2020

\bibitem{Caselle:2019khe}
M.~Caselle, A.~Nada, M.~Panero and D.~Vadacchino,
``Conformal field theory and the hot phase of three-dimensional U(1) gauge theory,''
JHEP \textbf{05}, 068 (2019).
%doi:10.1007/JHEP05(2019)068
%[arXiv:1903.00491 [cond-mat.str-el]].
%2 citations counted in INSPIRE as of 09 Jun 2020

\bibitem{ref:review:1}
P. Mehta, M. Bukov, C. Wang, A. Day, C. Richardson, C. Fisher, D. Schwab,
“A high-bias, low-variance introduction to Machine Learning for physicists,”
Phys. Rep. {\bf 810}, 1 (2019).
%	10.1016/j.physrep.2019.03.001
%arXiv:1803.08823.

\bibitem{ref:review:2}
G.~Carleo, I.~Cirac, K.~Cranmer, L.~Daudet, M.~Schuld, N.~Tishby, L.~Vogt-Maranto and L.~Zdeborová,
``Machine Learning and the Physical Sciences'',
Rev. Mod. Phys. {\bf 91}, 045002 (2019).
%10.1103/RevModPhys.91.045002
%arXiv:1903.10563.

\bibitem{ref:phases:0}
K. Zhou, G. Endrodi, L.-G. Pang, and H. Stöcker,
``Regressive and generative neural networks for scalar field theory,''
Phys. Rev. D 100, no. 1, 011501, (2019).
%http://arxiv.org/abs/1810.12879
%10.1103/PhysRevD.100.011501

\bibitem{ref:phases:1}
K. Ch’ng, J. Carrasquilla, R. G. Melko, and E. Khatami,
``Machine Learning Phases of Strongly Correlated Fermions,''
Phys. Rev. X {\bf 7}, 031038 (2017).

\bibitem{ref:phases:2}
J. Carrasquilla and R. G. Melko,
``Machine learning phases of matter,''
Nature Physics {\bf 13}, 431 (2017).

\bibitem{ref:phases:3}
E. P. L. van Nieuwenburg, Y.-H. Liu, and S. D. Huber,
``Learning phase transitions by confusion,''
Nature Physics {\bf 13}, 435 (2017).

\bibitem{ref:phases:4}
J. Venderley, V. Khemani, and E.-A. Kim,
``Machine Learning Out-of-Equilibrium Phases of Matter,''
Phys. Rev. Lett. {\bf 120}, 257204 (2018).

\bibitem{ref:phases:6}
Y.-H. Liu and E. P. L. van Nieuwenburg, 
``Discriminative Cooperative Networks for Detecting Phase Transitions'', Phys. Rev. Lett. \textbf{120}, 176401 (2018).

\bibitem{ref:phases:7}
R. A. Vargas-Hernández, J. Sous, M. Berciu, and R. V. Krems, 
``Extrapolating quantum observables with machine learning: Inferring multiple phase transitions from properties of a single phase'', 
Phys. Rev. Lett. \textbf{121}, 255702 (2018).

\bibitem{ref:phases:8}
B. S. Rem, N. Kaming, M. Tarnowski, L. Asteria, N. Fläschner, C. Becker, K. Sengstock and C. Weitenberg, 
``Identifying Quantum Phase Transitions using Artificial Neural Networks on Experimental Data'', 
Nat. Phys. \textbf{15}, 917 (2019).


\bibitem{ref:phases:9}
P. Broecker, J. Carrasquilla, R. G. Melko, and S. Trebst, 
“Machine learning quantum phases of matter beyond the fermion sign problem”, Scientific Reports \textbf{7}, 8823 (2017).

\bibitem{ref:phases:10}
Y.~Abe, K.~Fukushima, Y.~Hidaka, H.~Matsueda, K.~Murase and S.~Sasaki,
``Image-processing the topological charge density in the $\mathbb C P^{N-1}$ model,''
PTEP \textbf{2020}, no.1, 013D02 (2020).
%doi:10.1093/ptep/ptz134
%[arXiv:1805.11058 [hep-lat]].

\bibitem{ref:phases:11}
S. J. Wetzel, ``Unsupervised learning of phase transitions: from principal component analysis to variational autoencoders'', 
Phys. Rev. E \textbf{96}, 022140 (2017).

\bibitem{ref:phases:12}
S.~J.~Wetzel and M.~Scherzer,
``Machine Learning of Explicit Order Parameters: From the Ising Model to SU(2) Lattice Gauge Theory,''
Phys. Rev. B \textbf{96}, no.18, 184410 (2017).
%doi:10.1103/PhysRevB.96.184410
%[arXiv:1705.05582 [cond-mat.stat-mech]].

\bibitem{Bachtis:2020dmf}
D.~Bachtis, G.~Aarts and B.~Lucini,
``Extending Machine Learning Classification Capabilities with Histogram Reweighting,''
[arXiv:2004.14341 [cond-mat.stat-mech]].
%0 citations counted in INSPIRE as of 19 Jun 2020


\bibitem{Berezinskii:1971fst}
V.~L.~Berezinskii, 
``Destruction of Long-range Order in One-dimensional and Two-dimensional Systems Having a Continuous Symmetry Group I. Classical Systems'', Sov. Phys. JETP {\bf 32}, 493 (1970).

\bibitem{Berezinskii:1971scn}
``Destruction of Long-range Order in One-dimensional and Two-dimensional Systems Having a Continuous Symmetry Group II. Quantum Systems'' Sov. Phys. JETP {\bf 34}, 610 (1971).

\bibitem{Kosterlitz:1973xp}
J.~Kosterlitz and D.~Thouless,
``Ordering, metastability and phase transitions in two-dimensional systems,''
J. Phys. C \textbf{6}, 1181-1203 (1973)
%doi:10.1088/0022-3719/6/7/010

\bibitem{Dunne:2000vp}
G.~V.~Dunne, I.~I.~Kogan, A.~Kovner and B.~Tekin,
``Deconfining phase transition in (2+1)-dimensions: The Georgi-Glashow model,''
JHEP \textbf{01}, 032 (2001)
%doi:10.1088/1126-6708/2001/01/032
%[arXiv:hep-th/0010201 [hep-th]].

\bibitem{ref:Gattringer}
C. Gattringer, C.B. Lang, ``Quantum Chromodynamics on the Lattice'' (Springer-Verlag, Berlin Heidelberg, 2010).

\bibitem{ref:Omelyan}
I. P. Omelyan, I. M. Mryglod, and R. Folk,
``Optimized Verlet-like algorithms for molecular dynamics simulations'', Phys. Rev. E {\bf 65}, 056706 (2002) [cond-mat/0110438 [cond-mat.stat-mech]];
%I. P. Omelyan, I. M. Mryglod, and R. Folk,
``Symplectic analytically integrable decomposition algorithms: classification, derivation, and application to molecular dynamics, quantum and celestial mechanics simulations'', Comput. Phys. Commun. {\bf 151}, 272 (2003).


\bibitem{ref:book-ml:1}
  I.~Goodfellow, Y.~Bengio and A.~Courville,
  ``Deep Learning''
  (The MIT Press, Cambridge, Massachusetts, 2016).

\bibitem{ref:book-ml:2}
  F. Chollet,
  ``Deep Learning with Python''
  (Manning Publications, Shelter Island, New York, 2017).

\bibitem{ref:book-ml:3}
  A. Géron,
  ``Hands-On Machine Learning with Scikit-Learn, Keras, and TensorFlow''
  (2nd edition, O'Reilly Media, 2019).


\end{thebibliography}
\end{document}